\documentclass[11pt]{article}
\usepackage{graphicx}

\begin{document}

\def\xslash#1{{\rlap{$#1$}/}}
\def \p {\partial}
\def \dd {\psi_{u\bar dg}}
\def \ddp {\psi_{u\bar dgg}}
\def \pq {\psi_{u\bar d\bar uu}}
\def \jpsi {J/\psi}
\def \psip {\psi^\prime}
\def \to {\rightarrow}
\def\bfsig{\mbox{\boldmath$\sigma$}}
\def\DT{\mbox{\boldmath$\Delta_T $}}
\def\xit{\mbox{\boldmath$\xi_\perp $}}
\def \jpsi {J/\psi}
\def\bfej{\mbox{\boldmath$\varepsilon$}}
\def \t {\tilde}
\def\epn {\varepsilon}
\def \up {\uparrow}
\def \dn {\downarrow}
\def \da {\dagger}
\def \pn3 {\phi_{u\bar d g}}

\def \p4n {\phi_{u\bar d gg}}

\def \bx {\bar x}
\def \by {\bar y}

\begin{center}
{\Large\bf  QCD Factorization for Quarkonium Production in Hadron Collions at Low Transverse Momentum}
\par\vskip20pt
J.P. Ma$^{1,2}$ and C. Wang$^{3}$      \\
{\small {\it
$^1$ Institute of Theoretical Physics, Academia Sinica,
P.O. Box 2735,
Beijing 100190, China\\
$^2$ Center for High-Energy Physics, Peking University, Beijing 100871, China  \\
$^3$ School of Physics, Peking University, Beijing 100871, China  
}} \\
\end{center} 
\vskip 1cm
\begin{abstract}
Inclusive production of a quarkonium $\eta_{c,b}$ in hadron collisions at low transverse momentum can be used to extract various Transverse-Momentum-Dependent(TMD) gluon distributions of hadrons, provided the TMD factorization for the process
holds.  The factorization involving unpolarized TMD gluon distributions of unpolarized hadrons 
has been examined with on-shell gluons at one-loop level. In this work 
we study the factorization at one-loop level with diagram approach in the most general case, where all TMD gluon distributions  at leading twist are involved. We find that the factorization holds and the perturbative effects are represented by one perturbative coefficient. Since the initial gluons from hadrons are off-shell in general, there exists the so-called super-leading region found recently. We find that the contributions from this region can come from individual diagrams at one-loop level, but they are cancelled in the sum. Our factorized result for the differential cross-section is explicitly gauge-invariant.

\vskip 5mm
\noindent
\end{abstract}  
\vskip 1cm
\par 

\par\vskip20pt

\par
\noindent 
{\bf 1. Introduction} 
\par 
Theoretical predictions of the inclusive quarkonium production at large transverse momentum in hadron collisions can be
made by using the standard collinear factorization of QCD. Using the predictions one can extract from experimental results 
the gluon distribution functions of initial hadrons. These distribution functions are important for making predictions of other processes and for providing  the information about inner structure of initial hadrons. However the information is limited 
because the extracted gluon distributions are one-dimensional. It is possible to extract three-dimensional gluon distributions, called as Transverse-Momentum-Dependent(TMD) gluon distributions,   
by using processes involving small transverse momenta. For this purpose,  one needs to establish TMD factorizations. 

\par       
In TMD factorization nonperturbative- and perturbative 
effects in a process are consistently separated. TMD parton distributions are defined with QCD operators and represent 
the separated nonperturbative effects.   TMD quark distributions can be extracted from 
processes of Drell-Yan and Semi-Inclusive DIS. For these two processes, TMD factorizations
have been established in \cite{CSS,JMYP} and in  \cite{JMY,CAM}. It is noted that TMD factorization has been first established   
for inclusive $e^+e^-$-annihilations into two nearly back-to-back hadrons in the seminal work in \cite{CSEE}, where 
only TMD parton fragmentation functions are involved. For extracting TMD gluon distributions   
it is suggested  to use processes in hadron collisions like Higgs-production\cite{JMYG,SecG1}, quarkonium production \cite{BoPi}, two-photon production\cite{QSW}, 
the production of a quarkonium combined with a photon\cite{DLPS} and double-quarkonium 
production\cite{GPZ}. In \cite{SecG2,SecG3} the TMD distribution of linearly polarized gluons 
inside a nucleus and its phenomenology have been studied. 
\par 
TMD factorization for the suggested processes has been derived only at tree-level, except 
the process of Higgs- and quarkonium production. At one-loop level, 
TMD factorization has been examined for Higgs-production in \cite{JMYG} and 
for the production of $\eta_{c,b}$ in \cite{MWZ}. In these studies one takes the incoming 
gluons from the initial hadrons as on-shell and spin-averaged. The corresponding perturbative 
coefficient is determined at one-loop. Since only on-shell gluons with spin averaged 
are considered,   one examines in fact the TMD factorization  for the contribution from the scattering of unpolarized gluons 
coming from unpolarized hadrons. By taking the transverse momenta of incoming gluons into 
account, the gluon as a parton from an unpolarized hadron can be linearly polarized according to \cite{TMDGP}.  Certainly one can use on-shell-gluon scattering as in the studies of \cite{JMYG,MWZ} to study the factorization 
of the contribution from linearly polarized gluons at one-loop. But this is difficult because one needs to study the on-shell gluon scattering at three-loop, as discussed in \cite{MWZ}.  The factorization 
for the  processes with polarized hadrons has not been examined beyond tree-level. 

\par 

We will use the subtractive approach to examine TMD factorization for the inclusive production of a $^1S_0$-quarkonium $\eta_Q$, which is $\eta_c$ or $\eta_b$. The approach
is based on diagram expansion at hadron level and explained in \cite{JC1,JC2,JC3}.   In this approach the nonperturbative effects can be systematically subtracted in terms of diagrams. Using the approach TMD factorization for Drell-Yan processes has been examined\cite{MaZh}, where TMD quark distributions are involved.  In the case with TMD gluon distributions the situation is rather subtle, e.g., there can be super-leading-power contributions related to gluons as found in \cite{JC1}, and TMD factorization can be violated in certain processes as shown in \cite{JCJQ,JCV,TCR}.  
We notice here that gluons from initial hadrons participating in a hard scattering are in general off-shell, i.e., the gluons have momenta slightly off-shell and unphysical polarizations.  This brings up complications in examining 
the factorization and for obtaining gauge invariant results. The complications do not appear 
in \cite{JMYG,MWZ}.  In the subtractive approach the complications can be correctly addressed. 
\par   
In our work, we take the initial hadrons as arbitrarily polarized. We can show that 
at one-loop level the factorization holds and is explicitly gauge invariant.   The contributions from the super-leading region do not appear at tree-level in our case, but they do appear at one-loop level. 
The nonzero contributions come from two different sets of diagrams. They are canceled in the sum. 
Our result shows that  
there is only one perturbative coefficient for all various 
contributions involving different TMD gluon distributions. The coefficient 
is determined at one-loop.  
\par 
The above discussions are mainly relevant to the initial hadrons. For the nonperturbative effects related to the quarkonium 
in the final state, one can employ the factorization with nonrelativistic QCD(NRQCD) proposed in \cite{nrqcd}. A quarkonium 
can be taken as a bound state of a heavy quark $Q$ and a heavy anti-quark $\bar Q$. The heavy quark or heavy anti-quark 
moves with a small velocity $v$ in the rest frame of the quarkonium. For the production of $\eta_Q$, the production rate at the leading order of $v$ can be
written as a product of the production rate of a $Q\bar Q$ pair in color- and spin singlet with a NRQCD matrix element.
The NRQCD matrix element characterizes the transmission of the produced 
$Q\bar Q$ into $\eta_Q$.  In this work we will also add the
correction at the next-to-leading order of $v$ in NRQCD factorization at tree-level, since the correction and 
the one-loop correction can be of the same importance.  We notice here that in NRQCD factorization 
for a $P$-wave quarkonium one needs to consider not only the color-singlet $Q\bar Q$ pair, but also the color-octet $Q\bar Q$ pair\cite{nrqcd}. Taking the color-octet $Q\bar Q$ pair into account, TMD factorization for the production of a $P$-wave 
quarkonium is violated\cite{MWZ2}. It is noted that NRQCD factorization with the color-octet $Q\bar Q$ pair is also violated 
at two-loop level. But this violation can be avoided by adding gauge links in NRQCD color-octet matrix elements\cite{NQS}.   

\par 
Our work is organized as in the following: In Sect.2. we study TMD factorization at tree-level. Notations 
are introduced.  In Sect.3. we study TMD factorization at one-loop level. We will show that one needs to introduce a soft 
factor to complete TMD factorization. The mentioned complications from unphysical incoming gluons 
will be explained in detail. 
In Sect.4. we will give our main result of TMD 
factorization for the differential cross-section with arbitrary hadrons in the initial state. The differential cross-section is given in detail 
in the case that one of the initial hadrons is unpolarized and another one is of spin-1/2. Sect.5. is our summary.

\par\vskip20pt
\noindent 
{\bf 2. TMD Factorization at Tree-Level}
\par 
In this section we first introduce notations and TMD gluon density matrix or distributions in Subsection 2.1. 
In Subsection 2.2. we consider the tree-level contribution with one-gluon exchange. The contribution from two-gluon 
exchange is studied in Subsection 2.3. 

\par\vskip10pt
\noindent
{\bf 2.1. Notations and TMD Gluon Distributions}

\par   
For our purpose it is convenient to use the  light-cone coordinate system, in which a
vector $a^\mu$ is expressed as $a^\mu = (a^+, a^-, \vec a_\perp) =
((a^0+a^3)/\sqrt{2}, (a^0-a^3)/\sqrt{2}, a^1, a^2)$ and $a_\perp^2
=(a^1)^2+(a^2)^2$. We introduce two light-cone vectors:  $n$ and $l$. With these light-cone vectors one can 
build two tensors in the transverse space. The vectors and tensors are:  
\begin{equation}
n^\mu = (0,1,0,0), \quad l^\mu =(1,0,0,0), \quad  g_\perp^{\mu\nu} = g^{\mu\nu} - n^\mu l^\nu - n^\nu l^\mu, 
\quad  
\epsilon_\perp^{\alpha\beta} =  \epsilon^{\mu\nu\alpha\beta} l_\mu n_\nu, 
\end{equation} 
with $\epsilon_\perp^{12}=-\epsilon_\perp^{21}=1$. 
\par 
We consider the process 
\begin{equation} 
   h_A (P_A) + h_B (P_B) \to \eta_Q (q) + X,
\label{proc}    
\end{equation} 
where $\eta_Q$ stands for $\eta_c$ or $\eta_b$. It is a $^1 S_0$-quarkonium consisting of a heavy $Q\bar Q$ pair with $Q=c$ or $Q=b$.  
The momenta in Eq.(\ref{proc}) are  given by:
\begin{equation} 
  P_A^\mu \approx (P_A^+,0,0,0), \quad  P_B^\mu \approx (0,P_B^-, 0,0), \quad q^\mu =(q^+,q^-,\vec q_\perp)= (xP_A^+, y P_B^-,  \vec q_\perp),  
\end{equation}      
where we have neglected the masses of hadrons, i.e., $P_A^- \approx 0$ and $P_B^+ \approx 0$. 
The mass of the quarkonium is $M_\eta$ and the invariant mass of the initial hadrons is $s \approx 2 P_A^+ P_B^-$. At the leading order of the velocity expansion one has 
$M_\eta \approx 2 M_Q$. 
 We will also use the notation $q^2=Q^2=M^2_\eta$. We are interested in the kinematical region 
 of $q_\perp/ Q \ll 1$. In this region, one can establish TMD factorization to express the differential cross-section 
 in terms of TMD gluon distributions. 

\par 
TMD gluon distributions are defined with QCD operators. To define them for $h_A$ we introduce 
the gauge link along the direction $u^\mu =(u^+,u^-,0,0)$ with $u^+\ll u^-$: 
\begin{equation}
{\mathcal L}_u (z) = P \exp \left ( -i g_s \int^0_{-\infty}  d\lambda
     u\cdot A (\lambda u + z) \right ),   
\end{equation}
where the gluon field $A^\mu$ is in the adjoint representation. 
The gluonic density matrix of $h_A$ is defined as: 
 \begin{eqnarray}
 \Gamma_A ^{\mu\nu } (k ) =
\frac{1}{x P_A^+} \int \frac{ d\xi^- d^2 \xi_\perp}{(2\pi)^3}
e^{ - i \xi\cdot k }
 \langle h_A \vert \biggr ( G^{+\mu } (\xi ) {\mathcal L}_u (\xi) \biggr )^a
            \biggr ( {\mathcal L}_u^\dagger (0) G^{+\nu }(0) \biggr )^a  \vert h_A \rangle, 
\label{DEF} 
\end{eqnarray}
with $\xi^\mu =(0,\xi^-,\vec\xi_\perp)$ and $k^\mu = (x P_A^+, 0,  \vec k_\perp)$. $G^{\mu\nu}$ is the field strength tensor. The definition is gauge-invariant. The gauge links in Eq.(\ref{DEF}) are taken off light-cone. This is to avoid light-cone singularities if we take $u^+ =0$. In this work we will use Feynman gauge. In this non-singular gauge 
fields at infinite space-time are zero. If one works with a singular gauge, one needs to implement gauge links along the transverse direction at $\xi^- =-\infty$ to make the definition gauge-invariant\cite{GL1,GL2}.       
\par
The density matrix is defined for hadrons with an arbitrary spin. In general one can decompose 
the density matrix with scalar functions. These functions are TMD gluon distributions. 
Here we briefly discuss the case for $h_A$ with spin-1/2. We assume that the spin of $h_A$ is described by the helicity 
$S_L$ and the transverse spin $s_\perp^\mu$. At leading power or leading twist the indices $\mu$ and $\nu$ of the density matrix are transverse. The decomposition reads\cite{TMDGP}:
\begin{eqnarray}
 \Gamma_{A}^{\mu\nu}(k) &=& -\frac{1}{2} g_\perp^{\mu\nu} f_g (x,k_\perp)
   + \frac{1}{2 M_A^2} \left (k_\perp^\mu k_\perp^\nu  +  \frac{1}{2} g_\perp^{\mu\nu} k_\perp^2 \right ) H^\perp (x,k_\perp )
 + \frac{1}{2} S_L \biggr [ 
   -i\epsilon_\perp^{\mu\nu} \Delta G_L (x,k_\perp)  
\nonumber\\   
  &&  -\frac{1}{M_A^2}\tilde k_\perp^{\{\mu} k_\perp^{\nu\}}  \Delta H_L^\perp (x,k_\perp) \biggr ]  
 + \frac{1}{2 M_A } \biggr [  - g_\perp^{\mu\nu}  k_\perp\cdot \tilde s_\perp G_T (x, k_\perp) + i \epsilon_\perp^{\mu\nu} k_\perp\cdot s_\perp \Delta G_T (x,k_\perp) 
\nonumber\\
   &&  -\frac{1}{2 } \left ( \tilde k_\perp^{\{\mu} s_\perp^{\nu\}} +  \tilde s_\perp^{\{\mu} k_\perp^{\nu\}}  \right ) 
         \biggr ( \Delta H_T (x,k_\perp) - \frac{k_\perp^2}{2 M_A^2} \Delta H_T^{\perp} (x,k_\perp) \biggr ) 
\nonumber\\   
    &&  + \frac{1}{M_A^2}  \tilde k_\perp^{\{\mu} k_\perp^{\nu\}} k_\perp\cdot s_\perp \Delta H_T^\perp (x,k_\perp) \biggr ]  , 
\label{Dec-M} 
\end{eqnarray}
with $\tilde k_\perp^\mu =\epsilon_\perp^{\mu\nu} k_{\perp\nu}$ and $\tilde s_\perp^\mu =\epsilon_\perp^{\mu\nu} s_{\perp\nu}$.  
The TMD gluon distribution $f_g$ corresponds to the standard gluon distribution function of an unpolarized hadron 
if we integrate out the transverse momentum formally. By taking the transverse momentum into account, the gluon as a parton 
in an unpolarized hadron can be linearly polarized indicated by $H^{\perp}$ in Eq.(\ref{Dec-M}).

\par 
Similarly, one can define the gluon density matrix $ \Gamma_B^{\mu\nu} (k)$ or TMD gluon distributions of $h_B$, where one uses instead of ${\mathcal L}_{u}$ the gauge link ${\mathcal L}_{v}$ along the direction $v$ with $v^-\ll v^+$.  In the following 
subsections and Sect.3 we will set  $u=n$ and $v=l$  for convenience to study the factorization. It is well-known 
that there will be light-cone singularities in TMD gluon density matrices with the setting. But this will not affect 
our analysis, one can always make in each step the substitution $n\to u$ and $l\to v$ for subtractions. We will make the substitution in our final result.    

\par

\par 
\begin{figure}[hbt]
\begin{center}
\includegraphics[width=11cm]{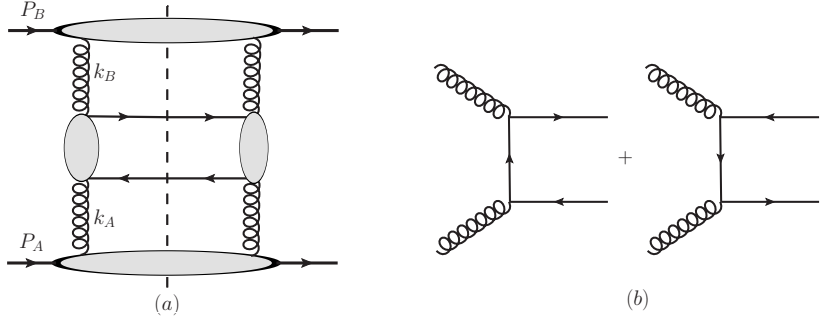}
\end{center}
\caption{(a): Tree-level diagram of one-gluon exchange. (b):  The bubbles in the middle of Fig.1a stand for the amplitude of $g^* g^* \to \eta_Q$.    }
\label{P1}
\end{figure}
\par

\par\vskip20pt
\noindent 
{\bf 2.2. TMD Factorization of One-Gluon-Exchange Contributions}
\par
At tree-level the contribution to the differential cross-section of Eq.(\ref{proc}) are from diagrams represented 
by Fig.1a. It can be written as: 
\begin{eqnarray} 
\frac{ d\sigma}{ d^4 q } 
  &=& \frac{1}{ 2s (2\pi)^3} \int d^4 k_A d^4 k_B \delta (q^2-4 m_Q^2) (2\pi)^4 \delta^4 (k_A + k_B-q) 
      {\mathcal M}^{ab}_{\alpha\mu} 
     (k_A,k_B)  \left ( {\mathcal M}^{cd}_{\beta\nu} 
     (k_A,k_B) \right ) ^\dagger    
\nonumber\\ 
   && \cdot \biggr [ \int \frac{d^4 \xi}{(2\pi)^4} 
   e^{i\xi\cdot k_A} \langle h_A \vert A^{c,\beta} (0) A^{a,\alpha}(\xi) \vert h_A \rangle \biggr ] 
      \cdot \biggr [ \int \frac{d^4 \eta}{(2\pi)^4} 
   e^{i\eta \cdot k_B} \langle h_B \vert A^{d,\nu} (0) A^{b,\mu}(\eta) \vert h_B \rangle \biggr ]. 
\label{Tree1}   
\end{eqnarray} 
In Eq.(\ref{Tree1}), the correlation function of gluon fields in the first $[\cdots]$ of the second line is represented 
by the lower bubble in Fig.1a, the correlation function in the second $[\cdots]$ is represented 
by the upper bubble. The bubble in the left-middle part of Fig.1a is the amplitude 
given by Fig.1b, i.e., $ {\mathcal M}^{ab}_{\alpha\mu} (k_A,k_B)$ is the amplitude of $g^*(k_A) g^* (k_B) \to \eta_Q$.
At amplitude level, there is only one gluon exchanged between bubbles. There can be more exchanged gluons. 
The case of two-gluon exchange will be studied in the next subsection.

\par 
We will take the leading order of the small velocity expansion in NRQCD\cite{nrqcd}. 
At the leading order, 
the quark $Q$ and $\bar Q$ carry the same momentum $q/2$ in Fig.1a. The broken line there is the cut. The cut cutting the lines of the heavy quark pair means also that we take the projection for the pair into the color-singlet $^1 S_0$-state. 
The projection is standard and can be found, e.g., in \cite{LMC}.  Under the above approximation, one easily finds the amplitude given in Fig.1b:
\begin{eqnarray} 
 {\mathcal M}^{ab,\alpha\mu}(k_A,k_B)   =  g_s^2 \delta^{ab} \epsilon^{\alpha\mu\sigma\rho}k_{A\sigma} k_{B\rho} f_L(k_A^2,k_B^2), \quad  \quad f_L(0,0)= -  
        \frac{\psi^*(0)}{m_Q^2 \sqrt{2 N_c m_Q}}, 
\label{TreeF}    
\end{eqnarray} 
where we introduced the form factor $f_L$ for two-gluon fusion into $\eta_Q$. The two gluons are in general off-shell. 
The explicit result of $f_L(0,0)$ for on-shell gluons is given. Here we use the wave function at the origin for the projection. 
In the final result it will be replaced with NRQCD matrix element. 

\par 
In the kinematical region under our consideration the produced $\eta_Q$ carries a small transverse momentum $q_\perp \sim \lambda Q$
with $\lambda\ll 1$. The leading power contributions arise when upper- and lower bubbles are jet-like functions. The power 
counting for the momenta carried by the gluons leaving the bubbles are: 
\begin{equation} 
   k_A^\mu \sim (1,\lambda^2, \lambda,\lambda), \quad k_B^\mu \sim (\lambda^2, 1, \lambda,\lambda). 
\end{equation} 
The power counting for the gauge vector field $A^\mu$ in the correlation functions represented by the lower- or upper bubble is  the same as for the momenta given in the above, respectively.        
Taking the tree-level result in Eq.(\ref{TreeF}) 
we have the leading order contribution in $\lambda$ for the combination appearing in Eq.(\ref{Tree1}):
\begin{eqnarray} 
 {\mathcal M}^{ab}_{\alpha\mu} 
     (k_A,k_B) A^{a,\alpha}(\xi ) A^{b,\mu}(\eta) &=& g_s^2 f_L(0,0) 
         \delta^{ab} \epsilon_\perp^{\alpha\mu} \left ( k_A^+ A^a_{\perp\alpha}(\xi) 
           -k_{A\perp\alpha} A^{a,+}(\xi) \right ) 
\nonumber\\           
        && \cdot   \left ( k_B^- A^b_{\perp \mu}(\eta)  - k_{B\perp\mu} A^{b,-}(\eta) \right ) + {\mathcal O}(\lambda^3) , 
\label{G2AMP}                    
\end{eqnarray}   
the leading order is at $\lambda^2$.  
In the expansion one also needs to expand $f_L(k_A^2,k_B^2)$ in $\lambda$. This gives the factor $f_L(0,0)$ in the above.
From the power counting, the so-called super-leading region can appear here in the amplitude 
if there is a contribution proportional to $ A^{a,+}(\xi ) A^{b,-}(\eta)$. Such a contribution can be at order of 
$\lambda^0$ or $\lambda^1$.  Usually, if we take the gluons from $h_{A,B}$ 
as on-shell, as the explicit calculation in \cite{MWZ}, one will not meet the super-leading 
regions, because the on-shell gluons are always transversely polarized.  In Eq.(\ref{G2AMP}) the contribution proportional to $ A^{a,+}(\xi ) A^{b,-}(\eta)$ is combined 
with two transverse momenta. Therefore, the super-leading region gives no contribution here.    We also note that the introduced form factor 
$f_L(k_A^2,k_B^2)$ in general is not gauge-invariant, but $f_L(0,0)$ is gauge-invariant obviously.
\par

Since we are only interested in the leading order of $\lambda$, the $-$-components of gauge fields in the correlation function of $h_A$ can always be neglected. The $-$-components of momenta carried by these fields can always be neglected except in the correlation function. The latter results in that one can perform the integration over those 
$-$-components of momenta and the corresponding $+$-components of the space-time coordinate vectors in the correlation 
function trivially. Therefore, we introduce the notations which will be used through this work: For any gauge field $A^\alpha(\tilde \xi)$ in the correlation function of $h_A$ the space-time coordinate vector is $\tilde\xi^\mu =(0, \xi^-, \vec\xi_\perp)$. The momentum $k$ carried by this field is $k^\mu =(k^+,0, \vec k_\perp)$. Similarly, we also use these notations for the correlation function of $h_B$ in which the role played by $+(-)$-components is exchanged by that of $-(+)$-components, 
respectively.

\par  

With the above result the one-gluon contribution from Fig.1a at the leading power of $\lambda$ is:
\begin{eqnarray} 
\frac{ d\sigma}{ d^4 q} 
  &=& \frac{\pi \delta (xy s-Q^2) }{ s} g_s^4 f_L^2 (0,0) \int d^2 k_{A\perp} d^2 k_{B\perp} \delta^2 (k_{A\perp} + k_{B\perp} -q_\perp)  
 \delta^{ab}\delta^{cd} \epsilon_{\perp\alpha\mu}\epsilon_{\perp\beta\nu}   
\nonumber\\ 
  && \cdot  \biggr [\int \frac{d^3 \tilde \xi}{(2\pi)^3} 
   e^{i\tilde \xi\cdot k_A} \langle h_A \vert \hat G^{c,+\beta} (0) \hat G^{a, +\alpha}(\tilde \xi) \vert h_A \rangle  \biggr ]
 \cdot \biggr [  
   \int \frac{d^3 \tilde \eta }{(2\pi)^3} 
   e^{i\tilde \eta \cdot k_B} \langle h_B \vert \hat G^{d,-\nu} (0) \hat G^{b,-\mu}(\tilde \eta) \vert h_B \rangle \biggr ],        
\label{1G}    
\end{eqnarray} 
with $k_A^-=0$ and $k_B^+=0$. Approximately one may write the two correlation functions in the second line as the gluon density matrix $\Gamma_A$ and $\Gamma_B$,
respectively. However, in these correlation functions, $\hat G^{\mu\nu}$ given by $\hat G^{\mu\nu}=\partial^\mu G^\nu -\partial^\nu G^\mu$ is not exactly      
the field strength tensor operator. To obtain it, one needs to consider the contributions from the exchange of two- or more gluons     

\par 
\begin{figure}[hbt]
\begin{center}
\includegraphics[width=15cm]{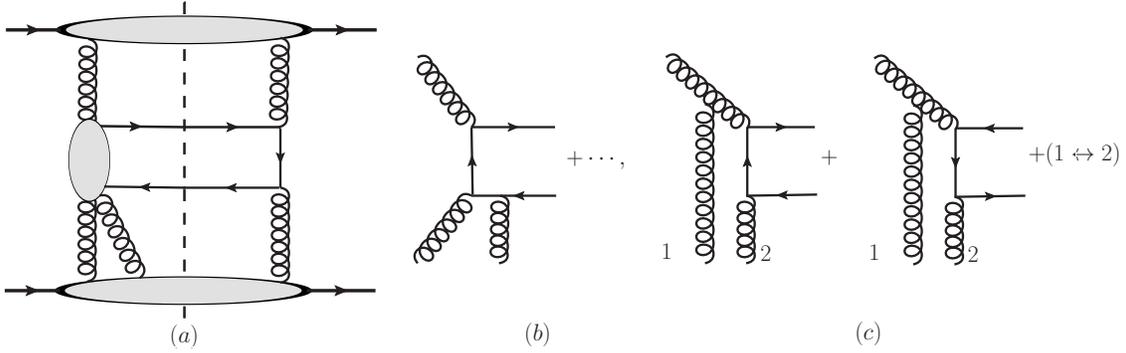}
\end{center}
\caption{ (a) The diagram with two-gluon change. The middle bubble in the 
left part is the sum of diagrams given in (b) and (c).  (b) The tree amplitude for $g^* g^* g^* \to \eta_Q$, in which all gluons are attached to the quark line.  (c) The tree amplitude for $g^* g^* g^* \to \eta_Q$, in which a three-gluon vertex is involved.  
  }
\label{P2}
\end{figure}
\par

\par\vskip10pt
\noindent 
{\bf 2.3. TMD Factorization of Two-Gluon-Exchange Contributions}
\par 

Now we consider the contribution in which two gluons come from $h_A$ as in those diagrams given in Fig.2a. 
The contribution after neglecting the unimportant components of momenta and with the notation introduced before Eq.(\ref{1G})  is: 
\begin{eqnarray} 
\frac{ d\sigma}{ d^4 q } 
  &=& \frac{\pi }{ s} \int d^3 k_1 d^3 k_2 d^3 k_{B}  \delta (q^2- 4m_Q^2) \delta^4(k_1 +k_2 + k_{B} -q) 
  {\mathcal M}^{a_1a_2 b}_{\alpha_1 \alpha_2 \mu} 
     (k_1,k_2, k_B)   
\nonumber\\
   &&  \cdot  \biggr ( \frac{1}{2} \biggr ) \biggr [\int \frac{d^3 \tilde \xi_1 d^3\tilde \xi_2}{(2\pi)^6} 
   e^{i\tilde  \xi_1\cdot k_1 +i\tilde \xi_2 \cdot k_2} \langle h_A \vert A^{c,\beta} (0) A^{a_1,\alpha_1}(\tilde \xi_1) A^{a_2,\alpha_2}( \tilde \xi_2) \vert h_A \rangle 
   \biggr ]
\nonumber\\   
   && \cdot \left ( {\mathcal M}^{cd}_{\beta\nu} 
     (k_1+k_2,k_B) \right ) ^\dagger\biggr [  
   \int \frac{d^3 \tilde  \eta }{(2\pi)^3} 
   e^{i\tilde \eta\cdot   k_B} \langle h_B \vert A^{d,\nu} (0) A^{b,\mu}(\tilde \eta) \vert h_B \rangle
     \biggr ],                
\label{TwoG1}     
\end{eqnarray}
The amplitude ${\mathcal M}^{a_1a_2 b}_{\alpha_1 \alpha_2 \mu} (k_1,k_2, k_B) $ is the sum of contributions 
from Fig.2b and Fig.2c. It is the amplitude for three-gluon fusion into $\eta_Q$ with the exclusion of those diagrams 
in which the gluon $1$ and gluon $2$ combine into one gluon with the three-gluon vertex. Those diagrams are in fact included 
in the lower bubble. A factor $1/2$ has to be implemented to avoid double-counting.

\par 
The three-gluon amplitude given by Fig.2b and Fig.2c will be used extensively in our work. Hence, we give the detailed 
result for the expansion with the power counting in $\lambda$. In the expansion one should keep  
the $i\varepsilon$ factors in propagators, because these factors have a physical meaning as we will see. 
The calculation of diagrams are straightforward, although it is tedious. But after the expansion the result 
takes a rather simple form for combinations with gauge fields appearing  in Eq.(\ref{TwoG1}). We have from 
Fig.2b and Fig.2c the following: 
\begin{eqnarray}
  &&{\mathcal M}^{a_1a_2 b}_{\alpha_1 \alpha_2 \mu} 
     (k_1,k_2, k_B)  \biggr\vert_{2b} \int \frac{ d^3 \tilde \xi_1 d^3  \tilde\xi_2 d^3 \tilde \eta}{(2\pi)^9} e^{ + i \tilde\xi_1\cdot k_1 + i \tilde \xi_2\cdot k_2 
       + i \tilde \eta\cdot k_B }   A^{a_1,\alpha_1}( \tilde\xi_1) A^{a_2,\alpha_2}( \tilde\xi_2) A^{b,\mu}( \tilde\eta) 
\nonumber\\ 
 && \approx - i g_s^3 f^{a_1a_2b} f_L(0,0)   \int \frac{ d^3  \tilde\xi_1 d^3 \tilde \xi_2 d^3  \tilde\eta}{(2\pi)^9} e^{ + i \tilde\xi_1\cdot k_1 +  i \tilde \xi_2\cdot k_2 
       + i \tilde\eta\cdot k_B }  \epsilon_{\perp\alpha\mu}  \hat G^{b,-\mu}( \tilde\eta)
\nonumber\\
&& \quad \cdot \biggr [ -i A^{a_1,\alpha} (\tilde \xi_1)  A^{a_2, +}( \tilde \xi_2)   + i  A^{a_2,\alpha} (\tilde \xi_2)  A^{a_1, +}( \tilde \xi_1)
   - \hat G^{a_1,+\alpha} (\tilde \xi_1)  A^{a_2, +}(\tilde \xi_2) \frac{1}{k_1^+ - i\varepsilon}\nonumber\\
 && \quad    + 
 \hat G^{a_2,+\alpha} (\tilde \xi_2) A^{a_1, +}( \tilde \xi_1)  \frac{1}{k_2^+ - i\varepsilon}  \biggr ] 
   - ({\mathcal B^-}-{\rm term}) ,
\nonumber\\     
  &&{\mathcal M}^{a_1a_2 b}_{\alpha_1 \alpha_2 \mu} 
     (k_1,k_2, k_B)  \biggr\vert_{2c} \int \frac{ d^3 \tilde \xi_1 d^3  \tilde\xi_2 d^3 \tilde \eta}{(2\pi)^9} e^{ + i \tilde\xi_1\cdot k_1 + i \tilde \xi_2\cdot k_2 
       + i \tilde \eta\cdot k_B }   A^{a_1,\alpha_1}( \tilde\xi_1) A^{a_2,\alpha_2}( \tilde\xi_2) A^{b,\mu}( \tilde\eta) 
\nonumber\\
&& \approx  -i g_s^3 f^{a_1a_2b} f_L(0,0)  \int \frac{ d^3 \tilde \xi_1 d^3\tilde \xi_2 d^3 \tilde \eta}{(2\pi)^9} e^{ + i\tilde\xi_1\cdot k_1 + i\tilde\xi_2\cdot k_2 
       + i\tilde\eta\cdot k_b }  \epsilon_{\perp\alpha\mu}  \hat G^{b,-\mu}(\tilde\eta)   \biggr [ 
        \biggr ( \frac{1}{k_1^+ -i\varepsilon} +\frac{1}{k_2^+ + i\varepsilon} \biggr )   
\nonumber\\       
 &&  \quad  \hat G^{a_1,+\alpha} (\tilde \xi_1)  A^{a_2, +}(\tilde \xi_2) - \biggr ( \frac{1}{k_1^+ + i\varepsilon} +\frac{1}{k_2^+ - i\varepsilon} \biggr )  \hat G^{a_2,+\alpha} (\tilde \xi_2)  A^{a_1, +}(\tilde \xi_1) \biggr ] 
   + ({\mathcal B^-}-{\rm term}),
 \label{GAC}                      
 \end{eqnarray} 
where the $   ({\mathcal B^-}-{\rm term})$ is proportional to $A^{b,-}(\tilde \eta)$.  It is exactly cancelled in the sum of Fig.2b and Fig.2c. The above results are symmetric in exchange of the gluon $1$ and $2$. Taking the sum we obtain the two-gluon-exchange contribution from Fig.2a at the leading order of $\lambda$:
\begin{eqnarray} 
\frac{ d\sigma}{ d^4 q } 
  &=& \frac{\pi \delta (xy s-Q^2) }{ s}  g_s^5 f_L^2 (0,0) \int d^3 k_1 d^3 k_{B} \delta^4(k_1  + k_{B} -q) 
        \epsilon_{\perp\alpha\mu}\left (  - \delta^{cd} \epsilon_{\perp\beta\nu}   \right )   
\nonumber\\
   && \biggr [  
   \int \frac{d^3 \tilde \eta }{(2\pi)^3} 
   e^{i \tilde \eta\cdot  k_B} \langle h_B \vert \hat G^{d,-\nu} (0) \hat G^{b,\mu}(\tilde \eta) \vert h_B \rangle
     \biggr ] \cdot  \biggr \{ \int \frac{d^3\tilde \xi_1 }{(2\pi)^3} 
   e^{i\tilde \xi_1\cdot k_1 } \langle h_A \vert \hat G^{c,+\beta} (0) f^{a_1a_2 b} 
\nonumber\\    
  && \biggr [  A^{a_1,+}(\tilde \xi_1) A^{a_2,\alpha}(\tilde \xi_1) 
 - i  \int d^3 k_2 \frac{ d^3\tilde  \xi_2}{(2\pi)^3} e^{ik_2 \cdot ( \tilde \xi_2-\tilde \xi_1)} 
        \frac{1}{k_2^+ + i\varepsilon} A^{a_2,+}(\tilde \xi_2) \hat G^{a_1,+\alpha}(\tilde \xi_1) \biggr ]     \vert h_A \rangle 
   \biggr \} .
\label{2G}     
\end{eqnarray}
We note here that the $\pm i\epsilon$'s in the denominators come from different places. 
The factor $+i\varepsilon$ comes from the gluon propagators in Fig.2c and indicates that the interactions through the exchange 
of the collinear gluon are of the initial-state, while the factor $-i\varepsilon$ comes from the quark propagators in Fig.2b 
and Fig.2c and indicates final-state interactions. In the sum the effects of final-state interactions are completely 
canceled and there is no term with the eikonal propagator $1/(k_{1,2}^+ - i\varepsilon)$. 
The sum only contains those eikonal propagators $1/(k_{1,2}^+ + i\varepsilon)$ representing initial-state interactions. 
This result is important here for the factorization. If the effects of initial-state- and final-state interactions 
exist simultaneously, they can be potential sources for breaking the factorization, as discussed in \cite{JCJQ,JCV}, or one needs to introduce non-universal TMD gluon distributions for the factorization\cite{NUN}. Then, the prediction power is lost.         
  
Now we  consider the product in the gluon density matrix $\Gamma_A^{\mu\nu}$ defined in Eq.(\ref{DEF}):  
\begin{eqnarray} 
  \biggr ( {\mathcal L}_n^\dagger (\tilde \xi) G^{+\mu}( \tilde \xi) \biggr )^a &=& \hat G^{a, +\mu}( \tilde \xi) -g_s f^{abc} A^{b,+}(\tilde \xi) 
     A^{c,\mu} ( \tilde \xi) 
\nonumber\\     
    &&  -ig_s f^{abc} \int d^3 k \frac{1}{k^+ + i\varepsilon} \int \frac{d^3  \tilde \xi_2}{(2\pi)^3} 
        e^{i k\cdot (\tilde \xi_2-\tilde \xi)} A^{b,+}(\tilde \xi_2) \hat G^{c,+\mu} (\tilde \xi) + {\mathcal O}(g_s^2). 
\end{eqnarray}
Comparing this expression with Eq.(\ref{2G}), one can realize that the first term in Eq.(\ref{2G}) is the missing part for the field strength tensor $G^{+\alpha}$ in the one-gluon-exchange contribution, and the second term forms 
a part of the gauge link. 
One may consider the exchange of more gluons to obtain full gauge links. Therefore, we can write 
the result of TMD factorization at tree-level as:            
\begin{eqnarray} 
\frac{ d\sigma}{ d^4 q} 
  &=& \frac{xy \pi \delta (xy s-Q^2) }{2 (N_c^2-1) } g_s^4 f_L^2 (0,0) \int d^2 k_{A\perp} d^2 k_{B\perp} \delta^2 (k_{A\perp} + k_{B\perp} -q_\perp)  
  \epsilon_{\perp\alpha\mu}\epsilon_{\perp\beta\nu}   
\nonumber\\ 
  &&  \Gamma_A^{\beta\alpha}(k_A) \Gamma_B^{\nu\mu}(k_B) \biggr ( 1 + {\mathcal O}(\alpha_s) + {\mathcal O} (\lambda) \biggr ) 
\label{TMDTREE}       
\end{eqnarray}
with $k_A^\mu = (xP_A^+,0,\vec k_{A\perp})$ and $k_B^\mu =(0, yP_B^-,\vec k_{B\perp})$.  In the above, the gluon density matrices 
are defined gauge-invariantly. $f_L(0,0)$ is gauge invariant as discussed before. Hence, the result in Eq.(\ref{TMDTREE}) 
is gauge invariant. This result can be still represented by Fig.1a, where the lower bubble represents $\Gamma_A^{\beta\alpha}$, the upper bubble represents $\Gamma_{B}^{\nu\mu}$. The obtained result in Eq.(\ref{TMDTREE}) will be corrected beyond tree-level.    

\par\vskip10pt

\par\vskip20pt 
\noindent 
{\bf 3. TMD Factorization at One-Loop-Level}
\par 
One-loop correction comes from diagrams in which one has one-gluon-exchange in the middle part of Fig.1. 
The one-loop correction can be divided into two parts. One part is the real part, in which an additional gluon is exchanged in the middle part of Fig.1a crossing the cut.  Another part is the virtual part, in which the exchanged gluon does not cross 
the cut. We will consider the two parts in the following subsections separately. 
\par 
\begin{figure}[hbt]
\begin{center}
\includegraphics[width=14cm]{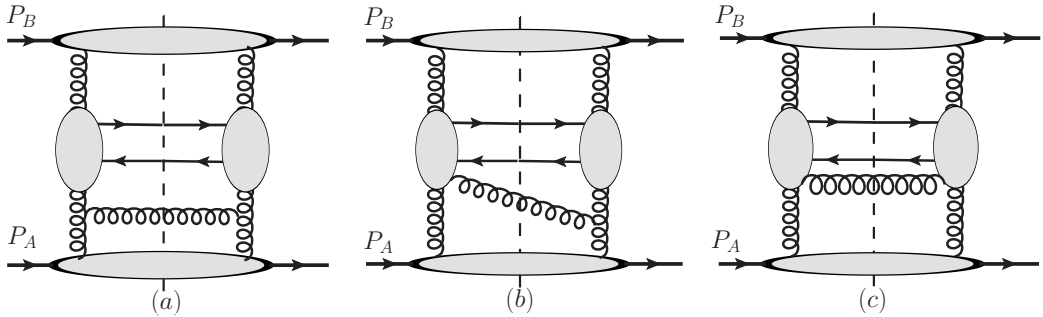}
\end{center}
\caption{Diagrams of the real part of one-loop correction.   }
\label{1LR}
\end{figure}
\par

\par\vskip10pt 
\noindent 
{\bf 3.1. The Real Part}
\par
The real part is given by the diagrams given in Fig.\ref{1LR}, where the bubbles in the middle 
represent the three-gluon- and two-gluon amplitudes explained in the last section. For the bubbles attached with three gluons, we always identify 
the two gluons coming from the below to the bubbles in the middle part  are the gluon $1$ and the gluon $2$, as specified in Fig.2c. With this identification, 
The total contribution of the real part is:
\begin{equation} 
  \frac{ d \sigma}{ d^4 q}\biggr\vert_R  =   \frac{ d \sigma}{ d^4 q}\biggr\vert_{\ref{1LR}a} 
       + \biggr ( \frac{ d \sigma}{ d^4 q}\biggr\vert_{\ref{1LR}b}    + h.c. \biggr ) + \frac{ d \sigma}{ d^4 q}\biggr\vert_{\ref{1LR}c}.          
\end{equation}       
In the kinematical region of $q_\perp \sim Q\lambda$ with $\lambda \ll 1$ 
the contributions at the leading order of $\lambda$ come from the region of the gluon momentum  which is collinear to $P_A$, or to $P_B$, and soft, with 
the standard power counting. We first consider the case that the exchanged gluon is collinear to $P_A$ in each diagram. 
\par 
The leading contribution from Fig.{\ref{1LR}a can be easily derived by using the leading result of the two-gluon amplitude in 
Eq.(\ref{G2AMP}) and by taking the pattern of the momentum $k$ of the exchanged gluon as $k^\mu \sim (1,\lambda^2,\lambda,\lambda)$. The leading contribution from Fig.\ref{1LR}a is: 
\begin{eqnarray} 
\frac{ d\sigma}{ d^4 q }\biggr\vert_{\ref{1LR}a}  
  &=& \frac{\pi \delta (xy s- Q^2) }{ s(N_c^2-1)} g_s^4 f_L^2 (0,0) \int d^3 k_1  d^3 k_{B}   \frac{ d^4 k}{(2\pi)^4}\delta^4(k_1  + k_{B} -q -k ) 
  (-2\pi \delta (k^2) )  \Gamma_B^{\nu\mu} (k_B)      
\nonumber\\   
   &&  
k_B^- \delta^{b d} \biggr ( (k_1-k)^+ g^{\sigma_1}_{\ \ \alpha_1}  -n_{\alpha_1} (k_1-k)_\perp^{\sigma_1} \biggr )       
  \biggr ( (k_1-k)^+ g^{\sigma}_{\ \ \beta_1} -n_{\beta_1} (k_1-k)_\perp^\sigma \biggr )
\nonumber\\
   &&\frac{- i} {(k_1-k)^2 + i\varepsilon} (-g_s f^{a b_1 b} )  \biggr  ( (-k-k_1)^{\alpha_1}  g^{\alpha}_{\  \rho} + 
    (2 k_1-k)_\rho g^{\alpha\alpha_1} + ( 2k -k_1)^{\alpha} g^{\alpha_1 }_{\ \rho}  \biggr )
\nonumber\\  
   && \frac{i} {(k_1-k)^2-i\varepsilon}  g_s f^{a c_1 d}  \biggr  (   (k+k_1)^{\beta_1} g^{\rho\beta} 
      + (-2 k_1+k)^\rho g^{\beta\beta_1} + (k_1-2k)^\beta g^{\rho\beta_1}  \biggr  )   
\nonumber\\
  &&  \epsilon_{\perp\sigma_1\mu} \epsilon_{\perp\sigma\nu}  \int\frac{d^3 \tilde \xi}{(2\pi)^3} e^{i \tilde \xi\cdot k_1} \langle h_A \vert A^{c_1}_\beta (0) 
      A^{b_1}_\alpha (\tilde \xi) \vert h_A \rangle   + \cdots .
\label{F3A}                         
\end{eqnarray}  
 In the above we have in fact included those contributions from exchange of all possible gluons between 
 the upper bubble and the bubbles in the middle of Fig.\ref{1LR}a. Therefore we have now in the above $\Gamma_B^{\nu\mu}$ 
 representing the upper bubble. This will be implicitly implied in the whole analysis. The $\cdots$ stand for power-suppressed 
 contributions which are neglected. 
 
 \par 
The power of $g_s$ indicates that the contribution in Eq.(\ref{F3A}) can be a part of the ${\mathcal O}(\alpha_s)$-correction in the tree-level factorization in Eq.(\ref{TMDTREE}). However, this needs to be examined. We note here that 
the $\Gamma_A^{\mu\nu}$ represented by the lower bubble in Fig.1a is a jet-like correlation function, which is the sum of all diagrams in which parton lines carry the momenta collinear to $P_A$.  
Because the exchanged gluon in Fig.\ref{1LR}a is collinear 
to $P_A$, the contribution from the exchanged gluon can be already included in the lower bubble $\Gamma_A^{\mu\nu}$ 
entirely or partly.  Therefore, the correct contribution to the ${\mathcal O}(\alpha_s)$-correction is only obtained 
after subtracting the corresponding contribution from $\Gamma_A^{\mu\nu}$. If one simply takes the contribution from Fig.3 as 
the ${\mathcal O}(\alpha_s)$-correction in Eq.(\ref{TMDTREE}) without the subtraction, a double-counting happens.  

\par 

\begin{figure}[hbt]
\begin{center}
\includegraphics[width=16cm]{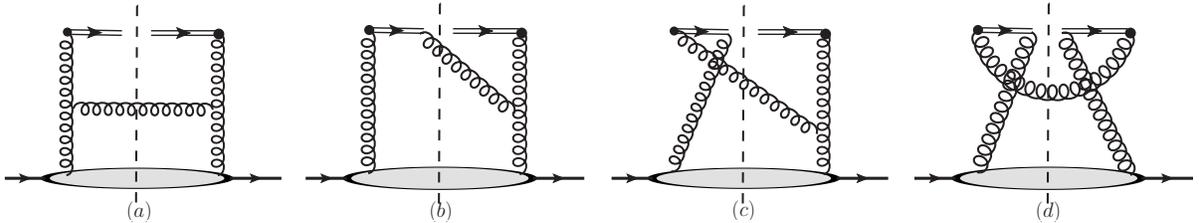}
\end{center}
\caption{Diagrams of the gluon TMD distribution for the subtraction}  
\label{1LRTMD}
\end{figure}
\par

\par 
 Beyond tree-level, the TMD gluon density matrix receives the contributions from diagrams in Fig.\ref{1LRTMD}. In these diagrams, the double lines represent the gauge links. The contribution from Fig.\ref{1LRTMD}a is:
\begin{eqnarray} 
\Gamma_A^{\mu\nu} ( k_A) \biggr\vert_{\ref{1LRTMD}a} &=& \frac{1}{xP^+} \int \frac{d^4 k d^3 k_1}{(2\pi)^4} (-2\pi \delta (k^2) )
     \delta^3 ( k_A - k_1 + k) (i) (k_A\cdot n g^{\nu\alpha_1} - (k_1-k)^\nu n^{\alpha_1})  
\nonumber\\
  &&\frac{- i} {(k_1-k)^2 + i\varepsilon} (-g_s f^{a b_1 c} ) ( (-k-k_1)^{\alpha_1}  g^{\alpha}_{\  \rho} + 
    (2 k_1-k)_\rho g^{\alpha\alpha_1} + ( 2k -k_1)^{\alpha} g^{\alpha_1 }_{\ \rho}  )
\nonumber\\
  && (-i) ( k_A\cdot n g^{\mu}_{\ \ \beta_1}  - (k_1-k)^\mu n_{\beta_1} )  
\nonumber\\  
   && \frac{i} {(k_1-k)^2-i\varepsilon} g_s f^{a c_1 c} \biggr  (   (k+k_1)^{\beta_1} g^{\rho\beta} 
      + (-2 k_1+k)^\rho g^{\beta\beta_1} + (k_1-2k)^\beta g^{\rho\beta_1} \biggr  )   
\nonumber\\
  && \int\frac{d^3 \tilde \xi}{(2\pi)^3} e^{i\tilde \xi\cdot k_1} \langle h_A \vert A^{c_1}_\beta (0) 
      A^{b_1}_\alpha (\tilde \xi) \vert h_A \rangle.
\label{F4A}          
\end{eqnarray}  
Comparing Eq.(\ref{F3A}) with Eq.(\ref{F4A}), we find that the contribution from Fig.\ref{1LR}a takes a factorized 
form: 
\begin{eqnarray} 
\frac{ d\sigma}{ d^4 q }\biggr\vert_{\ref{1LR}a}  
  &=& \frac{ x y \pi \delta (xy s-Q^2) }{ 2 (N_c^2-1) } g_s^4 f_L^2 (0,0) \int d^2 k_{A\perp}  d^2 k_{B\perp} \delta^2 (k_{A\perp}  + k_{B\perp} -q_\perp )       
\nonumber\\   
   &&  
 \epsilon_{\perp\alpha\mu} \epsilon_{\perp\beta\nu}       
     \Gamma_B^{\nu\mu} (k_B)   \biggr ( \Gamma_A^{\beta\alpha} (k_A)\biggr\vert_{\ref{1LRTMD}a} \biggr ).  
\end{eqnarray}  
This result indicates that the contribution from Fig.\ref{1LR}a is already included in the gluon density matrix 
$\Gamma_A^{\beta\alpha}$ in Eq.(\ref{TMDTREE}). Therefore, it should be subtracted 
from the ${\mathcal O}(\alpha_s)$-correction. This leads to that the contribution from Fig.\ref{1LR}a 
will not contribute to the ${\mathcal O}(\alpha_s)$-correction.        

\par 
Now we consider the contribution from Fig.\ref{1LR}b. Taking the exchanged gluon as collinear to $P_A$ and  the leading result for the two-gluon amplitude 
represented by the right bubble in the middle, the contribution reads:
\begin{eqnarray} 
\frac{ d\sigma}{ d^4 q }\biggr\vert_{\ref{1LR}b}  
  &= & \frac{\pi \delta (xy s-Q^2) }{ s} \int d^3 k_1  d^3 k_{B}   \frac{ d^4 k}{(2\pi)^4} \delta^4(k_1  + k_{B} -q -k ) 
  (-2\pi \delta (k^2) )       
\nonumber\\
   &&   {\mathcal M}^{a_1a_2 b}_{\alpha_1 \alpha_2 \mu}(k_1,-k, k_B)\biggr [\int \frac{d^3\xi}{(2\pi)^3} 
   e^{i\tilde \xi\cdot k_1 } \langle h_A \vert A^{c}_\beta (0) A^{a_1,\alpha_1}(\tilde \xi) \vert h_A \rangle 
   \biggr ]\biggr (-i g_s^3  f_L(0,0) \delta^{c_1d}  \biggr )  f^{ca_2c_1}      
\nonumber\\   
   &&   \biggr \{    \biggr  (  \epsilon_{\perp\beta_1\nu} (k_1-k)^+ 
        - n_{\beta_1} \epsilon_{\perp\sigma\nu} (k_1-k)_\perp^\sigma  \biggr )    
   \int \frac{d^3 \tilde \eta }{(2\pi)^3} 
   e^{i\tilde \eta\cdot  \tilde k_B} \langle h_B \vert  \hat G^{d,- \nu} (0) A^{b,\mu}(\tilde \eta) \vert h_B \rangle
     \biggr ]   
\nonumber\\
   &&  \biggr ( (-k_1-k)^{\beta_1} g^{\alpha_2 \beta} + (2k-k_1)^{\beta} g^{\alpha_2\beta_1} 
       + (2 k_1 -k)^{\alpha_2} g^{\beta\beta_1} \biggr ) \biggr \}  \frac{i}{(k_1-k)^2 -i\varepsilon}+\cdots.
\label{F4B}         
\end{eqnarray} 
In the above there are still some contributions at higher order of $\lambda$. These contributions need to be separated 
and neglected as those represented by $\cdots$. We note here that among diagrams in Fig.\ref{1LR}, the contribution from Fig.\ref{1LR}b is the most difficult to analyze. 

\par  
To find the leading contribution we can use the leading order result of the three-gluon amplitude in Eq.(\ref{GAC}). 
We expand the relevant combination in $\lambda$:   
\begin{eqnarray} 
  &&{\mathcal M}^{a_1a_2 b}_{\alpha_1 \alpha_2 \mu} 
     (k_1,-k, k_B) \int \frac{ d^3 \tilde \xi d^3 \tilde \eta}{(2\pi)^6} e^{  i\tilde \xi\cdot k_1  
       + i \tilde \eta\cdot k_B }   A^{a_1,\alpha_1}(\tilde \xi)  A^{b,\mu}(\tilde \eta) 
\nonumber\\
&& =  -g_s^3 f^{a_1a_2b}\frac{2 A_0}{m_Q^2}   \int \frac{ d^3 \tilde \xi d^3 \tilde \eta}{(2\pi)^6} e^{  i\tilde \xi_\cdot k_1  
       + i\tilde \eta\cdot k_B } \epsilon_{\perp}^{\sigma\rho} \hat G^{b,-}_{\quad\rho} (\tilde \eta)\biggr \{ \frac{1}{ k^+ -i\varepsilon} n_{\alpha_2}  
       \biggr (  (k_1^+-k^+) A^{a_1}_\sigma (\tilde \xi) - k_{1\perp\sigma} A^{a_1,+}(\tilde \xi)  \biggr )  
\nonumber\\              
     &&  +    \frac{1}{ k_1^+ + i\varepsilon} A^{a_1,+}( \tilde \xi) \biggr ( g_{\perp\alpha_2 \sigma} (k_1^+ -k^+) + n_{\alpha_2} k_{\perp\sigma} \biggr )  + l_{\alpha_2} ({\mathcal O(\lambda) ) + {\mathcal O}(\lambda^2} ) \biggr \},                           
\label{G44A}
\end{eqnarray}   
where the gluon field $A^{b,\mu}(\tilde\eta)$ is from the correlation function of $h_B$, and 
$A^{a_1,\alpha_1}(\tilde \xi_1)$ is from that of $h_A$. 
The combination is a vector with the index $\alpha_2$ which will be contracted with the remaining terms in Eq.(\ref{F4B}). 
This index is carried by the exchanged gluon. It is noted that the power for different $\alpha_2$ is different. The terms 
proportional to $n_{\alpha_2}$ in $\{\cdots\}$ are at order of $\lambda$, while the term with $\alpha_2 =\perp$ is at 
order of $\lambda^0$. At first look one may neglect the terms proportional to $n_{\alpha_2}$. But one can not neglect 
them, because all terms contracted with the remaining terms in Eq.(\ref{F4B}) will give the leading order contribution 
to the differential cross-section, except the terms proportional to $l_{\alpha_2}$.   

\par 
It will be lengthy to give the full result for the contribution from Fig.\ref{1LR}b. However, one can rather easily find
the factorized form of the contribution. We consider the contributions from Fig.\ref{1LRTMD}b and Fig.\ref{1LRTMD}c 
to the TMD gluon density matrix: 
\begin{eqnarray} 
 \Gamma_A^{\mu\nu} (k_A) \biggr\vert_{\ref{1LRTMD}b} &=& \frac{1}{xP^+} \int \frac{d^4 k d^3 k_1}{(2\pi)^4} (-2\pi \delta (k^2) )
     \delta^3 ( k_A -k_1 + k) (i) (k_A\cdot n g^{\nu\alpha} - k_1^\nu n^{\alpha})  
\nonumber\\
  &&  \frac{- i}{ n\cdot k -i\varepsilon} ( g_s  n_\rho f^{a c b_1} ) (-i) ( k_A\cdot n g^{\mu}_{\ \ \beta_1}  - (k_1-k)^\mu n_{\beta_1} )  
\nonumber\\  
   && \frac{i} {(k_1-k)^2-i\varepsilon} g_s f^{a c_1 c} \biggr (  (k+k_1)^{\beta_1} g^{\rho\beta} 
      + (-2 k_1+k)^\rho g^{\beta\beta_1} + (k_1-2k)^\beta g^{\rho\beta_1} \biggr  )   
\nonumber\\
  && \int\frac{d^3 \tilde \xi}{(2\pi)^3} e^{i \tilde \xi\cdot k_1} \langle h_A \vert A^{c_1}_\beta (0) 
      A^{b_1}_\alpha ( \tilde \xi) \vert h_A \rangle, 
\nonumber\\
 \Gamma_A^{\mu\nu} (k_A) \biggr\vert_{\ref{1LRTMD}c} &=& \frac{1}{xP^+} \int \frac{d^4 k d^3 k_1}{(2\pi)^4} (-2\pi \delta (k^2) )
     \delta^3 (k_A - k_1 + k) (i) (k_A\cdot n g^{\nu}_{\ \ \rho} + k^\nu n_{\rho})  
\nonumber\\
  &&  \frac{i}{  n\cdot k_1 + i\varepsilon} ( g_s  n^\alpha f^{b_1 a c} ) (-i) ( k_A\cdot n g^{\mu}_{\ \ \beta_1}  - (k_1-k)^\mu n_{\beta_1} )  
\nonumber\\  
   && \frac{i} {(k_1-k)^2-i\varepsilon} g_s f^{a c_1 c} \biggr (  (k+k_1)^{\beta_1} g^{\rho\beta} 
      + (-2 k_1+k)^\rho g^{\beta\beta_1} + (k_1-2k)^\beta g^{\rho\beta_1}  \biggr )   
\nonumber\\
  && \int\frac{d^3 \tilde \xi}{(2\pi)^3} e^{i\tilde \xi\cdot k_1} \langle h_A \vert A^{c_1}_\beta (0) 
      A^{b_1}_\alpha (\xi) \vert h_A \rangle.  
\end{eqnarray}        
With these expressions and result in Eq.(\ref{G44A}), we find that the contribution from Fig.\ref{1LR}b takes 
the factorized form:
\begin{eqnarray} 
\frac{ d\sigma}{ d^4 q }\biggr\vert_{\ref{1LR}b}  
  &=& \frac{ x y \pi \delta (xy s- Q^2) }{ 2 (N_c^2-1) } g_s^4 f_L^2 (0,0) \int d^2 k_{A\perp}  d^2 k_{B\perp}\delta^2(k_{A\perp}  + k_{B\perp} -q_\perp )       
\nonumber\\   
   &&  
 \epsilon_{\perp\alpha\mu} \epsilon_{\perp\beta\nu}       
     \Gamma_B^{\nu\mu} (k_B)   \biggr ( \Gamma_A^{\beta\alpha} (k_A)\biggr\vert_{\ref{1LRTMD}b}  + \Gamma_A^{\beta\alpha} (k_A)\biggr\vert_{\ref{1LRTMD}c} \biggr ). 
\label{F3BC}  
\end{eqnarray}  
In the above the contribution factorized with Fig.\ref{1LRTMD}b is from these terms proportional  to $1/(k^+- i\varepsilon)$ in the $\{\cdots\}$ 
in Eq.(\ref{G44A}), while the contribution factorized with Fig.\ref{1LRTMD}c is from these terms proportional to $1/(k^+_1+i\varepsilon) $.   
Again, the contribution from Fig.\ref{1LR}b is totally subtracted. It does not contribute to the ${\mathcal O}(\alpha_s)$-correction. 
\par 
The remaining diagram which needs to be studied is Fig.\ref{1LR}c. The contribution 
involves the three-gluon amplitudes only. The leading order result can be obtained  in a straightforward way by using the result in Eq.(\ref{G44A}).  The leading contribution 
comes from the term with the index $\alpha_2 =\perp$. We obtain the contribution at the leading order of $\lambda$:
\begin{eqnarray} 
\frac{ d\sigma}{ d^4 q }\biggr\vert_{\ref{1LR}c}  
  &=& \frac{\pi \delta (xy s- Q^2) }{ s (N_c^2-1)}  g_s^6 f_L^2 (0,0) \int d^3 k_1  d^3 k_{B} \frac{ d^4 k}{(2\pi)^4}  \delta^4(k_1  + k_{B} -q -k ) 
  (-2\pi \delta (k^2) )   \epsilon_{\perp\alpha\mu} \epsilon_{\perp\beta\nu}  k_B^-      
\nonumber\\
   &&    g_{\perp}^{\alpha\beta}  f^{a_1a_2b} f^{c_1a_2d} \biggr (\frac{(k_1-k) ^+}{k_1^+} \biggr )^2 \biggr [\int \frac{d^3\tilde \xi}{(2\pi)^3} 
   e^{i\tilde \xi\cdot k_1 } \langle h_A \vert A^{c_1,+} (0) A^{a_1,+}(\tilde \xi) \vert h_A \rangle 
   \biggr ]  \Gamma_B^{\nu\mu} (k_B)  
    +\cdots.  
\end{eqnarray} 
This contribution takes a factorized form with the contribution from Fig.\ref{1LRTMD}d 
to the TMD gluon density matrix:
\begin{eqnarray} 
  \Gamma_A^{\mu\nu} ( k_A) \biggr\vert_{\ref{1LRTMD}d} &=& \frac{1}{xP^+} \int \frac{d^4 k d^3 k_1}{(2\pi)^4} (-2\pi \delta (k^2) )
     \delta^3 ( k_A - k_1 + k) (i) (k_A\cdot n g^{\nu}_{\ \ \rho} + k^\nu n_{\rho}) g_s^2 f^{b_1 ac} f^{c_1 ac} 
\nonumber\\
  &&  (-i) ( k_A\cdot n g^{\mu\rho}  +k^\mu n^\rho )  \left (\frac{1}{n\cdot k_1 } \right )^2    
  \int\frac{d^3 \tilde \xi}{(2\pi)^3} e^{i \tilde \xi\cdot k_1} \langle h_A \vert A^{c_1,+}  (0) 
      A^{b_1,+} (\tilde \xi) \vert h_A \rangle.        
\end{eqnarray}
We have:
\begin{eqnarray} 
\frac{ d\sigma}{ d^4 q }\biggr\vert_{\ref{1LR}c}  
  &=& \frac{ x y \pi \delta (xy s- Q^2) }{ 2 (N_c^2-1) } g_s^4 f_L^2 (0,0) \int d^2 k_{A\perp}  d^2 k_{B\perp} \ \delta^2(k_{A\perp}  + k_{B\perp} -q_\perp )       
\nonumber\\   
   &&  
 \epsilon_{\perp\alpha\mu} \epsilon_{\perp\beta\nu}       
     \Gamma_B^{\nu\mu} (k_B)   \biggr ( \Gamma_A^{\beta\alpha} (k_A)\biggr\vert_{\ref{1LRTMD}d}   \biggr ).  
\end{eqnarray}  
This indicates that the contribution from Fig.\ref{1LR}c will not contribute to the ${\mathcal O}(\alpha_s)$-correction.        
\par 
From the above analysis, the leading contributions from the exchange of the gluon collinear to $P_A$ 
are already included in the gluon density matrix $\Gamma_A^{\mu\nu}$ in the factorized form in Eq.(\ref{TMDTREE}) 
at tree-level. Therefore, these leading contributions do not contribute to the ${\mathcal O}(\alpha_s)$-correction 
in Eq.(\ref{TMDTREE}). Performing a similar analysis one will also find that the leading contributions from the exchange of the gluon collinear to $P_B$ 
are already included in the gluon density matrix $\Gamma_B^{\mu\nu}$. 
We conclude here that the leading contributions 
of the real part from the exchange of a collinear gluon will not contribute to the ${\mathcal O}(\alpha_s)$-correction. 
They are correctly factorized into TMD gluon density matrices. Therefore,  the possible ${\mathcal O}(\alpha_s)$-correction to Eq.(\ref{TMDTREE})  can only come from the real part subtracted with the collinear contribution, i.e., from the difference:  
\begin{eqnarray}
  \frac{d\sigma }{ d^4 q} \biggr\vert_{R,s} &=& \frac{d\sigma }{ d^4 q} \biggr\vert_R  - \biggr [  \frac{ x y \pi \delta (xy s-Q^2) }{ 2 (N_c^2-1) } g_s^4 f_L^2 (0,0) \int d^2 k_{A\perp}  d^2 k_{B\perp} \delta^2 (k_{A\perp}  + k_{B\perp} -q_\perp )       
\nonumber\\   
   &&  
 \epsilon_{\perp\alpha\mu} \epsilon_{\perp\beta\nu}       
     \biggr (  \Gamma_B^{\nu\mu} (k_B)    \biggr (  \Gamma_A^{\beta\alpha} (k_A)  \biggr\vert_{Fig. \ref{1LRTMD}} \biggr )  
       + \Gamma_A^{\beta\alpha} (k_A) \biggr ( \Gamma_B^{\nu\mu} (k_B) \biggr\vert_{ Fig.\ref{1LRTMD}} \biggr )  \biggr )  + h.c. \biggr ] . 
\label{SG1}      
\end{eqnarray} 
The contribution from the above expression at the leading order of $\lambda$ can only come from the momentum 
region where the exchanged gluon is soft. We will study the soft-gluon contribution in the next subsection.     
\par 
\par 

Before turning to the soft-gluon contribution, it is interesting to compare the results here with the calculation done in \cite{MWZ} by replacing the initial hadrons with on-shell gluons.  In \cite{MWZ} one can only examine the factorization of  a part of the differential cross-section, in which only $f_g$ in Eq.(\ref{Dec-M}) is involved. i.e., the contribution from unpolarized gluons in  unpolarized 
hadrons. 
With the employed approach here, we are able to examine the factorization with the entire gluon density matrices. 
In this approach the initial gluons are in general off-shell. This brings up some additional complications in comparison 
with the study in \cite{MWZ}. The complications are indicated by the fact that we need to consider additional 
contributions represented by those contributions to gluon density matrices given by Fig.\ref{1LRTMD}c and Fig.\ref{1LRTMD}d.
\par

\par\vskip10pt
\noindent 
{\bf 3.2 Soft-Gluon Contributions and the Soft Factor}
\par 
In the kinematical region of $q_\perp/Q \sim \lambda \ll 1$, dominant contributions can come from the momentum region 
in which the exchanged gluon in Fig.\ref{1LR} is soft. The soft gluon carries the momentum $k$ at the order: 
\begin{equation} 
    k^\mu \sim (\lambda,\lambda,\lambda,\lambda). 
\end{equation}
In analyzing the collinear contributions, e.g., the three-gluon amplitude contracted with gauge fields in Eq.(\ref{GAC}), 
there are no contributions from the super-leading region in the set of diagrams given by Fig.2b or Fig.2c. 
However, we will have the contributions 
from the super-leading region in analyzing the contributions of soft gluons. We first consider the three-gluon contribution contracted with gauge fields as that in Eq.(\ref{G44A}). 
\par 
For the case of the soft gluon we have from Fig.2b and Fig.2c the contributions at the leading order of $\lambda$:     
\begin{eqnarray} 
  &&{\mathcal M}^{a_1a_2 b}_{\alpha_1 \alpha_2 \mu} 
     (k_1, -k , k_B)  \biggr\vert_{2c} \int \frac{ d^3 \tilde \xi_1 d^3 \tilde \eta}{(2\pi)^6} e^{ + i\tilde \xi_1\cdot k_1  
       + i \tilde \eta\cdot k_B }   A^{a_1,\alpha_1}( \tilde \xi_1)  A^{b,\mu}(\tilde \eta) 
\nonumber\\
&& = - g_s^3 f^{a_1a_2b} A_0  \int \frac{ d^3\tilde  \xi_1  d^3 \tilde \eta}{(2\pi)^6} e^{ + i \tilde \xi_1\cdot k_1 
       + i \tilde \eta\cdot k_B } \frac{4 i k_\perp^\beta } { q\cdot k -i\varepsilon }  \epsilon_{\perp\alpha_2 \beta}  A^{a_1,+}( \tilde \xi_1) 
          A^{b,-}( \tilde \eta) \biggr ( 1 +{\mathcal O}(\lambda) \biggr ), 
\nonumber\\ 
  &&{\mathcal M}^{a_1a_2 b}_{\alpha_1 \alpha_2 \mu} 
     (k_1, -k , k_B)  \biggr\vert_{2b} \int \frac{ d^3 \tilde \xi_1 d^3 \tilde \eta}{(2\pi)^6} e^{ + i\tilde \xi_1\cdot k_1  
       + i \tilde \eta\cdot k_B }   A^{a_1,\alpha_1}( \tilde \xi_1)  A^{b,\mu}(\tilde \eta) 
\nonumber\\
&& = g_s^3 f^{a_1a_2b} A_0  \int \frac{ d^3\tilde  \xi_1  d^3 \tilde \eta}{(2\pi)^6} e^{ + i \tilde \xi_1\cdot k_1 
       + i \tilde \eta\cdot k_B } \frac{4 i  k_\perp^\beta } { q\cdot k -i\varepsilon }  \epsilon_{\perp\alpha_2 \beta}  A^{a_1,+}( \tilde \xi_1) 
          A^{b,-}( \tilde \eta)   \biggr ( 1 +{\mathcal O}(\lambda) \biggr ).
\label{GSS1}                      
 \end{eqnarray} 
If we calculate the soft-gluon contributions to the differential cross-section by using the result from the set of diagrams in Fig. 2b or Fig.2c, one will find that each contribution will be at the order of $\lambda$ lower than that of the tree-level result given in Eq.(\ref{TMDTREE}).  This is the  
contribution from the super-leading region as discussed in \cite{JC1}.  However, the contributions from the super-leading region are canceled in the sum of Fig.2b and Fig.2c from Eq.(\ref{GSS1}).  
\par 
To find the contributions from the leading region, one has to expand the contribution from Fig.2b and Fig. 2c 
at the next-to-leading order of $\lambda$.  The contribution 
from each set of diagrams is very lengthy, but the sum takes the simple form: 
\begin{eqnarray} 
  &&{\mathcal M}^{a_1a_2 b}_{\alpha_1 \alpha_2 \mu} 
     (k_1,-k, k_B) \int \frac{ d^3 \tilde \xi_1 d^3 \tilde \eta}{(2\pi)^6} e^{ + i\tilde \xi_1\cdot k_1  
       + i \tilde \eta\cdot k_B }   A^{a_1,\alpha_1}( \tilde \xi_1)  A^{b,\mu}(\tilde \eta) 
\nonumber\\
&& =  - g_s^3 f^{a_1a_2b}A_0  \int \frac{ d^3 \tilde \xi d^3 \tilde \eta}{(2\pi)^6} e^{ + i\tilde \xi \cdot k_1  
       + i \tilde \eta\cdot k_B }  \frac{2}{m_Q^2 ( k^+ -i\varepsilon ) } \epsilon_{\perp}^{\sigma\rho} \hat G^{b,-}_{\quad\rho} (\tilde \eta) 
        \biggr [  g_{\sigma \alpha_2} k^+ A^{a_1,+} (\tilde \xi)    
\nonumber\\         
       && +n_{\alpha_2} \biggr (  k_1^+ A^{a_1}_\sigma (\tilde \xi) - k_{1\perp\sigma} A^{a_1,+}(\tilde \xi)  \biggr )  \biggr ] \biggr ( 1  +{\mathcal O}(\lambda) \biggr ) .                            
\label{GSG}
\end{eqnarray} 
Using this result we obtain the contribution from the soft-gluon exchange of Fig.\ref{1LR}b at the leading 
power:
\begin{eqnarray} 
\frac{ d\sigma}{ d^4 q }\biggr\vert_{\ref{1LR}b, s }  
  &=& \frac{\pi \delta (xy s- Q^2) }{ s} g_s^6 f_L^2 (0,0) \int d^3 k_A  d^3 k_{B} \int \frac{d^4 k}{(2\pi)^4} \delta^4(k_A  + k_{B} -q -k_\perp)        
 \epsilon_{\perp\sigma\rho} \epsilon_{\perp\mu\nu}          
\nonumber\\    
   &&   \Gamma_B^{\nu\rho } ( k_B)    \Gamma_A^{\mu\sigma} ( k_A) \frac{ k_B^- k_A^+}{N_c^2-1}  \biggr [ \frac{1}{N_c^2-1}  
      2\pi \delta (k^2) f^{abc} f^{abc} \frac{1}{k^+ -i\varepsilon } \frac{1}{k^-+i\varepsilon} \biggr ]  \biggr ( 1  +{\mathcal O}(\lambda) \biggr ), 
\label{F3BS}      
\end{eqnarray}
the factor in $[\cdots ]$ is a part of the soft factor introduced in the below. 
\par 
We note that there are nonzero soft-gluon contributions in the subtracted collinear contributions in Eq.(\ref{SG1}).  The contribution from the exchanged gluon collinear to $P_A$ from Fig.\ref{1LR}b given 
in Eq.(\ref{F3BC})  has the same soft-gluon contribution as given in Eq.(\ref{F3BS}).  After analyzing all 
contributions we have the difference in Eq.(\ref{SG1}) at the leading power: 
\begin{eqnarray} 
\frac{d\sigma}{ d^4 q}\biggr\vert_{R,s} &\approx& - \frac{\pi \delta (xy s- Q^2) }{ s} g_s^6 f_L^2 (0,0) \int d^3 k_A  d^3 k_{B} \int \frac{d^4 k}{(2\pi)^4} \delta^4(k_A  + k_{B} -q -k_\perp)        
 \epsilon_{\perp\sigma\rho} \epsilon_{\perp\mu\nu}          
\nonumber\\    
   &&   \Gamma_B^{\nu\rho } ( k_B)   \Gamma_A^{\mu\sigma} ( k_A) \frac{ k_B^- k_A^+}{N_c^2-1}  \biggr [ \frac{1}{N_c^2-1}  
      2\pi \delta (k^2) f^{abc} f^{abc} \frac{1}{k^+ -i\varepsilon } \frac{1}{k^-+i\varepsilon} \biggr ] + h.c. . 
\label{SG2}      
\end{eqnarray}
The difference is nonzero at the leading order of $\lambda$. 
If we take the factorization as given in Eq.(\ref{TMDTREE}), then the difference 
should be taken as the ${\mathcal O}(\alpha_s)$-correction. However, the difference is the effect 
of the soft gluon. It should be taken as a nonperturbative effect which needs to be factorized. 
For this one needs  to implement 
a soft factor in Eq.(\ref{TMDTREE}). 
   
\par 
The needed soft factor is defined as:  
\begin{equation} 
\tilde S(\vec\ell_\perp) = \int\frac{d^2 b_\perp}{(2\pi)^2} e^{ i\vec b_\perp\cdot \vec\ell_\perp}
  S^{-1}(\vec b_\perp)
\end{equation} 
with 
\begin{eqnarray}
S(\vec b_\perp) = \frac{1}{N_c^2-1} \langle 0\vert {\rm Tr} \left  [   {\mathcal L}^\dagger_v (\vec b_\perp,-\infty)
  {\mathcal L}_u (\vec b_\perp,-\infty) {\mathcal L}_u^\dagger (\vec 0,-\infty){\mathcal L}_v (\vec 0,-\infty)   \right ] \vert 0\rangle. 
\end{eqnarray}
This soft factor has been 
introduced in the TMD factorization of Higgs production in \cite{JMYG}.  At leading order one has
\begin{equation}
\tilde S(\vec\ell_\perp ) =\delta^2 (\ell_\perp) + {\mathcal O}(\alpha_s).
\end{equation}
\par
\begin{figure}[hbt]
\begin{center}
\includegraphics[width=10cm]{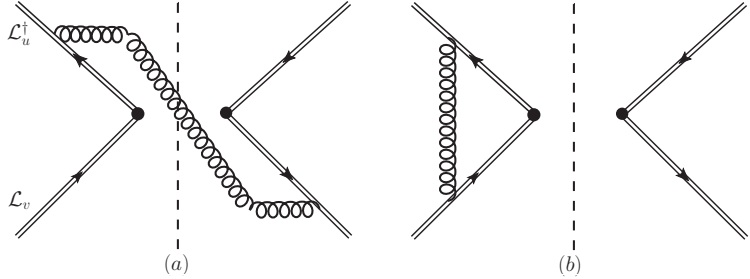}
\end{center}
\caption{One-loop corrections for the soft factor. The two diagrams with their conjugated diagrams give 
the one-loop correction for the soft factor with $u^+ =0$ and $v^-=0$. }
\label{soft}
\end{figure}
\par
For the convenience we take $u^+=0 $ and $v^-=0$ here. There are corrections at one-loop. 
One can divide the corrections into a virtual- and
a real part.  The real part is given by Fig.\ref{soft}a and the virtual part is given by Fig.\ref{soft}b. For $u^+\neq 0$ and $v^-\neq 0$ there are more diagrams in which one gluon is exchanged between gauge links along the same direction. We will 
come back later to the contributions from those diagrams.  The sum  of Fig.\ref{soft}a and its conjugated diagram reads:
\begin{eqnarray} 
\tilde S(\ell_\perp)\biggr\vert_{\ref{soft}a + h.c.  }  =   - g_s^2    \int \frac{d^4 k}{(2\pi)^4} \delta^2 (\ell_\perp -k_\perp)   \biggr [ \frac{1}{N_c^2-1}  
      2\pi \delta (k^2) f^{abc} f^{abc} \frac{1}{k^+ -i\varepsilon } \frac{1}{k^-+i\varepsilon} \biggr ] +h.c. .   
\end{eqnarray}
We note that the terms in $ [\cdots ]$ are exactly those in $[\cdots]$ of Eq.(\ref{SG2}).      
Now we modify the tree-level factorization in Eq.(\ref{TMDTREE}) as:
\begin{eqnarray} 
\frac{ d\sigma}{ d^4 q} 
   &=&  \frac{xy \pi \delta (xy s- Q^2) }{2 (N_c^2-1) } g_s^4 f_L^2 (0,0)  \int d^2 k_{A\perp} d^2 k_{B\perp} d^2 \ell_\perp \delta^2 (k_{A\perp} + k_{B\perp}+\ell_\perp  -q_\perp)
\nonumber\\    
   &&  \epsilon_{\perp\alpha\mu}\epsilon_{\perp\beta\nu}
        \Gamma_A^{\beta\alpha}( k_A) \Gamma_B^{\nu\mu} (k_B)  \tilde S(\ell_\perp) \biggr ( 1 + {\mathcal O}(\alpha_s) + {\mathcal O} (\lambda) \biggr ) .        
\label{FACS}    
\end{eqnarray} 
With this modification, one can see that the soft-gluon contribution in Eq.(\ref{SG2}) is now 
in the soft factor.  Combining the results in the last subsection we conclude that 
the real part of one-loop correction at the leading order of $\lambda$ are included 
in TMD gluon density matrices and the soft factor. The real part does not give 
the ${\mathcal O}(\alpha_s)$-correction. Only the virtual part can give the correction.

\par\vskip10pt

\begin{figure}[hbt]
\begin{center}
\includegraphics[width=15cm]{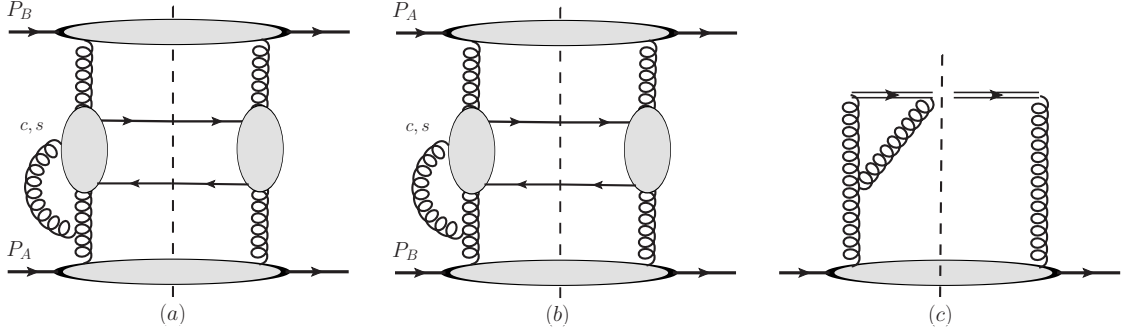}
\end{center}
\caption{Virtual corrections. }
\label{1LV}
\end{figure}
\par

\noindent 
{\bf 3.3. The Virtual Part} 
\par 
The virtual part receives contributions from two sets of diagrams. One set consists of diagrams, in which an additional gluon is exchanged  between quark lines  in diagrams given in Fig.1b. The contributions from this set of diagrams contain Coulomb- and I.R. divergences. These divergences 
are correctly factorized with NRQCD factorization into the NRQCD matrix element.  After NRQCD factorization, the contributions from this set will give a part of the ${\mathcal O}(\alpha_s)$ -correction.  Another set of diagrams consists of those in which an additional gluon is exchanged between a quark- and a gluon line or between gluon lines. This set of diagrams is given by Fig.{\ref{1LV}a and Fig.{\ref{1LV}b with the three-gluon amplitude defined before.  The virtual part can be written as the sum:   
\begin{equation}
   \frac{d\sigma}{d^4 q} \biggr\vert_V = \biggr ( \frac{d\sigma}{d^4 q}\biggr\vert_{\ref{1LV}a} + 
     \frac{d\sigma}{d^4 q}\biggr\vert_{\ref{1LV}b}   -\frac{d\sigma}{d^4 q}\biggr\vert_{G} + \frac{d\sigma}{d^4 q}\biggr\vert_{Q} \biggr )  + h.c., 
\end{equation} 
where the last term in $ (\cdots )$ is the contribution from the set of diagrams where an additional gluon 
is exchanged between quark lines.  In the sum of  first two terms, the contribution from the gluon-exchange between 
gluon lines is double-counted.  This is corrected by the third term which stands for the contribution from the gluon-exchange between gluon lines.  
As discussed, only the first three terms are  relevant to the TMD factorization.  

\par 
The contribution from Fig.\ref{1LV}a is given as:
\begin{eqnarray} 
\frac{ d\sigma}{ d^4 q }\biggr\vert_{\ref{1LV}a}  
  &=& \frac{\pi \delta (xy s- Q^2) }{ s} \int d^3 k_A  d^3 k_{B}\delta^4(k_A  + k_{B} -q  ) \int \frac{ d^4 k_2}{(2\pi)^4} 
    \frac{-i}{k_1^2+i\varepsilon}\frac{-i}{k_2^2+i\varepsilon}        
\nonumber\\
   &&   {\mathcal M}^{a_1a_2 b}_{\alpha_1 \alpha_2 \mu}(k_1,k_2, k_B)\biggr [\int \frac{d^3\xi}{(2\pi)^3} 
   e^{i\xi\cdot k_A } \langle h_A \vert A^{c}_\beta (0) A^{a}_{\alpha}(\xi) \vert h_A \rangle 
   \biggr ]  \left (\frac{1}{2}\right ) 
\nonumber\\   
   &&  \left ( {\mathcal M}^{cd}_{\beta \nu} 
     (k_A, k_B) \right ) ^\dagger\biggr [  
   \int \frac{d^3 \eta }{(2\pi)^3} 
   e^{i\eta\cdot  \tilde k_B} \langle h_B \vert A^{d,\nu} (0) A^{b,\mu}(\eta) \vert h_B \rangle
     \biggr ]
\nonumber\\
   && \biggr (- g_s f^{a_1 a_2 a} \biggr ) \biggr ( (-k_1+k_2)^\alpha g^{\alpha_1\alpha_2} + (-k_2-k_A)^{\alpha_1} g^{\alpha\alpha_2} 
       + (k_A+k_1)^{\alpha_2} g^{\alpha\alpha_1}  \biggr )   
\label{F6A}             
\end{eqnarray} 
with $k_1+k_2 =k_A$. A factor $1/2$ should be added to avoid  a double counting.  The contribution 
is essentially a part of one-loop correction to the introduced form factor $f_L(k_A^2,k_B^2)$. It contains 
 collinear- and infrared divergences. One may think that these divergences can be regularized 
 by the off-shellnesses $k_A^2$ and $k_B^2$.  However, in order to find the leading power contribution one has to expand the form factor 
in $k_A^2$ and $k_B^2$.  Only the contribution with $k_{A}^2 = k_B^2=0$  is at the leading power,  because of that the $\delta$-function $\delta^2(k_{A\perp}+k_{B\perp}-q_\perp)$ in Fig.\ref{1LV}a is at the oder of $\lambda^{-2}$. Therefore, 
the contribution from Fig.\ref{1LV}a at the leading power of $\lambda$ is a part of one-loop correction to the on-shell form factor $f_L(0,0)$.  
Similar situation also appears in TMD factorization of Drell-Yan processes and a detailed 
discussion about virtual corrections can be found in \cite{MaZh}. The collinear- and infrared divergences can be regularized with dimensional regularization. These divergences need to correctly be factorized or subtracted.  
\par 
 Using the early result one can obtain the collinear contribution in which the gluon with $k_2$ is collinear 
  to $P_A$:
\begin{eqnarray} 
\frac{ d\sigma}{ d^4 q }\biggr\vert_{\ref{1LV}a,c}  
  &=& \frac{\pi \delta (xy s- Q^2) }{ s}  g_s^5 f_L^2(0,0) \int d^3 k_A  d^3 k_{B}\delta^4(k_A  + k_{B} -q  ) \int \frac{ d^4 k_2}{(2\pi)^4} 
    \frac{-i}{k_1^2+i\varepsilon}\frac{-i}{k_2^2+i\varepsilon}        
\nonumber\\
   &&  f^{a_1a_2 b}  \epsilon_{\perp\rho\mu} 
      \frac{n_{\alpha_2 }}{k_2^+ + i\varepsilon  } \biggr ( k_A^+ g_\perp^{\rho\alpha_1} 
      - k_{1\perp}^\rho n^{\alpha_1}  \biggr )   
  \biggr [\int \frac{d^3\xi}{(2\pi)^3} 
   e^{i\xi\cdot k_A } \langle h_A \vert A^{c, \beta} (0) A^{a}_{\alpha}(\xi) \vert h_A \rangle 
   \biggr ]  
\nonumber\\   
   &&  (i  \delta^{cd}) \biggr (\epsilon_{\perp\beta\nu} k_A^+  
    - n_\beta  \epsilon_{\perp\sigma \nu} k_{A\perp}^\sigma \biggr  ) \biggr [  
   \int \frac{d^3 \eta }{(2\pi)^3} 
   e^{i\eta\cdot  \tilde k_B} \langle h_B \vert \hat G^{d,-\nu} (0) \hat G^{b, -\mu}(\eta) \vert h_B \rangle
     \biggr ]
\nonumber\\
   && \biggr (- g_s f^{a_1 a_2 a} \biggr ) \biggr ( (-k_1+k_2)^\alpha g^{\alpha_1\alpha_2} + (-k_2-k_A)^{\alpha_1} g^{\alpha\alpha_2} 
       + (k_A+k_1)^{\alpha_2} g^{\alpha\alpha_1}  \biggr ) . 
\label{F6AC}  
\end{eqnarray}
The corresponding contribution to the gluon TMD distribution from Fig.\ref{1LV}c is:
\begin{eqnarray} 
\Gamma_A^{\mu\nu} (k_A) \biggr\vert_{\ref{1LV}c} &=& \frac{1}{xP^+} \int \frac{d^4 k }{(2\pi)^4} \frac{-i}{k^2+i\varepsilon} 
      (i) (k_A\cdot n g^{\nu\alpha_1} - (k_A-k)^\nu n^{\alpha_1}) (-i) ( k_A\cdot n g^{\mu\beta}  - k_A ^\mu n^{\beta} ) 
\nonumber\\  
   && \frac{- i} {(k_A-k)^2 + i\varepsilon} (-g_s f^{d a b} ) (  (-k-k_A)^{\alpha_1} g^{\rho\alpha} 
      + (2 k_A  - k)^\rho g^{\alpha\alpha_1} + (-k_A + 2k)^\alpha g^{\rho\alpha_1} )   
\nonumber\\
  &&  \frac{ i}{ n\cdot k + i\varepsilon} ( g_s  n_\rho f^{d c b} )  \int\frac{d^3 \xi}{(2\pi)^3} e^{i\xi\cdot k_A} \langle h_A \vert A^{c}_\beta (0) 
      A^{a}_\alpha (\xi) \vert h_A \rangle, 
\label{F6C} 
\end{eqnarray}
where $k$ is the momentum of the gluon attached to the gauge link and is flowing into the gauge link.   
Comparing Eq.(\ref{F6AC}) with Eq.(\ref{F6C}), one finds: 
\begin{eqnarray} 
\frac{ d\sigma}{ d^4 q }\biggr\vert_{\ref{1LV}a,c}  
  &=&  \frac{ x y \pi \delta (xy s- Q^2) }{ 2 (N_c^2-1) } g_s^4 f_L^2 (0,0)  \int d^3 k_A  d^3 k_{B}\delta^4(k_A  + k_{B} -q  )
     \epsilon_{\perp\alpha\mu} \epsilon_{\perp\beta\nu} 
\nonumber\\
   && \Gamma_B^{\nu\mu} (k_B)   \left (\Gamma_A^{\beta\alpha}(k_A) \biggr\vert_{\ref{1LV}c} \right ), 
\end{eqnarray}  
therefore, the collinear contribution is already included in the TMD gluon density matrix.  Performing the analysis for the case that the gluon is collinear to $P_B$, one obtains the similar result. We then have the difference 
\begin{eqnarray}
  \frac{d\sigma }{ d^4 q} \biggr\vert_{V,s} &=& \frac{d\sigma }{ d^4 q} \biggr\vert_V  - \biggr [  \frac{ x y \pi \delta (xy s- Q^2) }{ 2 (N_c^2-1) } g_s^4 f_L^2 (0,0) \int d^2 k_{A\perp}  d^2 k_{B\perp} \ \delta^4(k_{A\perp}  + k_{B\perp} -q_\perp )       
\nonumber\\   
   &&  
 \epsilon_{\perp\alpha\mu} \epsilon_{\perp\beta\nu}       
     \biggr (  \Gamma_B^{\nu\mu} (k_B)     \Gamma_A^{\beta\alpha} (k_A)\biggr\vert_{Fig. \ref{1LV}c} \    + \Gamma_A^{\beta\alpha} (k_A) \Gamma_B^{\nu\mu} (k_B)  \biggr\vert_{ Fig.\ref{1LV}c } \biggr )  + h.c. \biggr ] , 
\label{SG6}      
\end{eqnarray}   
which does not contain any collinear divergence.  But, the difference contains infrared divergences from 
the soft-gluon exchange. 
\par 
For the soft-gluon exchange we consider the case that $k_2^\mu$ is soft, 
i.e., $k_2^\mu \sim (\lambda,\lambda,\lambda,\lambda)$.  In this case, $k_1$ is collinear. 
Therefore, in analyzing the soft contribution from Fig.\ref{1LV}a, the factor $1/2$ 
should be replaced with $1$ to obtain the correct result.  Using the result in the last subsection for the three-gluon amplitude, we have the soft-gluon contribution:  
\begin{eqnarray} 
\frac{ d\sigma}{ d^4 q }\biggr\vert_{\ref{1LV}a,s}  
  &=&  \frac{ x y \pi \delta (xy s- Q^2) }{ 2 (N_c^2-1) } g_s^4 f_L^2 (0,0)   \int d^3 k_A  d^3 k_{B}\delta^4(k_A  + k_{B} -q  )
    \epsilon_{\perp\rho\mu}\epsilon_{\perp\beta \nu} 
\nonumber\\
  && \Gamma_B^{\nu\mu} (k_B) \Gamma_A^{\beta\rho} (k_A) 
   \biggr [  -  \frac{g_s^2}{N_c^2-1}  f^{a_1a_2a} f^{a_1a_2 a} \int \frac{ d^4 k_2}{(2\pi)^4} 
   \frac{-i}{k_2^2+i\varepsilon} \cdot  \frac{i}{k _2^- - i\varepsilon}\cdot  \frac{i}{k_2^+ +  i\varepsilon}  \biggr ].       
\label{F6AS}  
\end{eqnarray}
Again,  there is also a soft-gluon contribution in the gluon TMD distribution from Fig.\ref{1LV}c with 
the same factor in $[\cdots ]$ in Eq.(\ref{F6AS}). 
At the end one finds the soft-gluon contribution for the difference in Eq.(\ref{SG6}) which is similar  to the case of real corrections:
\begin{eqnarray}
  \frac{d\sigma }{ d^4 q} \biggr\vert_{V,s} &=& - 
 \frac{ x y \pi \delta (xy s- Q^2) }{ 2 (N_c^2-1) } g_s^4 f_L^2 (0,0)   \int d^3 k_A  d^3 k_{B}\delta^4(k_A  + k_{B} -q  )
    \epsilon_{\perp\rho\mu}\epsilon_{\perp\beta\nu} \Gamma_B^{\nu\mu} (k_B)
\nonumber\\
  && \Gamma_A^{\beta\rho} (k_A) 
   \biggr [  -  \frac{g_s^2}{N_c^2-1}  f^{a_1a_2a} f^{a_1a_2 a} \int \frac{ d^4 k_2}{(2\pi)^4} 
   \frac{-i}{k_2^2+i\varepsilon} \cdot  \frac{i}{k _2^- - i\varepsilon}\cdot  \frac{i}{k_2^+ +  i\varepsilon}  \biggr ] 
   + h.c., 
\end{eqnarray}
This soft-gluon contribution is in fact included in the soft factor. The soft factor receives the contribution from Fig.\ref{soft}b and its conjugated diagram. It is:
\begin{eqnarray} 
\tilde S(\ell_\perp,\mu,\rho)\biggr\vert_{\ref{soft}b +h.c.} =   \delta^2 (\ell_\perp)  \frac{g_s^2}{N_c^2-1}  f^{abc} f^{abc} 
    \int \frac{d^4 k}{(2\pi)^4} \frac{-i}{k^2+i\varepsilon} \cdot   \frac{i} {k^-  - i\varepsilon} 
   \cdot  \frac{i}{k^+ + i\varepsilon} +h.c. .  
\end{eqnarray}   
If we take the factorized form as in Eq.(\ref{FACS}), then the  soft-gluon contribution is included in the soft factor. 
\par 
Based on the results in this subsection and previous ones, we find that at the leading power of $q_\perp\sim \lambda Q$ with $\lambda \ll 1$, the real part of the one-loop correction are correctly factorized into TMD gluon densities and the 
introduced soft factor. It will not contribute to the ${\mathcal O}(\alpha_s)$-correction. The virtual part gives 
contributions to the correction, but the collinear- and infrared divergences are subtracted into TMD gluon density matrices and the soft factor, respectively.  The correction is finite.  

\par\vskip20pt
\noindent 
{\bf 4. The Final Result} 
\par 
In the study of the previous sections, we have set $u=n$ and $v=l$ for convenience. This setting will
generate light-cone singularities in the subtraction. To present the final result, we undo the setting for TMD gluon density matrices 
and the soft factor.  For $u^+\neq 0$ and $v^-\neq 0$ the TMD density matrix $\Gamma_A^{\mu\nu}$ or
$\Gamma_B^{\mu\nu}$ will depend on an extra parameter $\zeta_u$ or $\zeta_v$, respectively. The soft 
factor contains the parameter $\rho$. These parameters are defined as:
\begin{equation}
 \zeta^2_u = \frac {2 u^-}{u^+} \left ( P^+_A \right )^2, \quad \zeta_v^2 = \frac {2 v^+ }{v^-} \left ( P^-_B  \right )^2, \quad \rho^2 =\frac{ u^- v^+}{u^+ v^-}. 
 \end{equation}
Our final factorized result can be written as: 
\begin{eqnarray} 
\frac{ d\sigma}{ dx d y d^2 q_\perp} 
   &=&  \frac{2 \pi \delta (xy s- Q^2) }{Q^2} \sigma_0  \int d^2 k_{A\perp} d^2 k_{B\perp} d^2 \ell_\perp \delta^2 (k_{A\perp} + k_{B\perp}+\ell_\perp  -q_\perp) \epsilon_{\perp\alpha\mu}\epsilon_{\perp\beta\nu}\nonumber\\    
   && 
      {\mathcal H} (\zeta_u,\zeta_v, \rho )  \biggr (  \Gamma_A^{\beta\alpha}( k_A,\zeta_u ) \Gamma_B^{\nu\mu} (k_B,\zeta_v)  \biggr )  \tilde S(\ell_\perp,\rho ),       
\label{FACSF}    
\end{eqnarray} 
where the dependence on the renormalization scale $\mu$ in each term in the last line 
is suppressed. $\sigma_0$ is given by 
\begin{equation} 
\sigma_0 = \frac{(4\pi\alpha_s)^2 }{N_c (N_c^2-1) m_Q} \vert \psi(0)\vert^2= \frac{(4\pi\alpha_s)^2 }{2N_c^2 (N_c^2-1) m_Q} \langle {\mathcal O} (^1 S_0^{[1]} )\rangle .         
\end{equation} 
Here, we have expressed the quantity $\vert \psi (0)\vert^2$  with the corresponding NRQCD matrix element 
$\langle {\mathcal O} (^1 S_0^{[1]} ) \rangle$. The definition of the matrix element can be found in \cite{nrqcd}.
\par 
As mentioned in the subsection 3.2. there are more diagrams for the one-loop correction of the soft-factor 
for the case of $u^+\neq 0$ and $v^-\neq 0$. In these diagrams there is one-gluon exchange 
between gauge links along the same direction. In this case the TMD gluon density matrices also receive one-loop 
contributions from one-gluon exchange between gauge links along the same direction. These contributions are exactly 
canceled in Eq.(\ref{FACS}) or Eq.(\ref{FACSF}) by those from the soft factor.     
\par 
In Eq.(\ref{FACSF})  ${\mathcal H}$ is the perturbative coefficient which  starts at the order of $\alpha_s^0$.
Because of that all one-loop real corrections are subtracted into TMD gluon density matrices and the soft factor from our analysis in Sect. 3.2., the coefficient is determined by the virtual correction. 
It is determined by the form factor of the fusion of two on-shell gluons into $\eta_Q$ after the subtraction of collinear- and infrared divergences with 
TMD gluon density matrices and the soft factor as shown in Sect. 3.3. Since there is only one form factor for the fusion, we have then correspondingly  in Eq.(\ref{FACSF}) only one perturbative coefficient. 
It is noted that the form factor of the fusion with two on-shell gluons and the subtraction are gauge-invariant,  ${\mathcal H}$  and Eq.(\ref{FACSF}) are hence also gauge-invariant.  
\par

At the leading order of $\alpha_s$  ${\mathcal H}$ is $1$. Beyond the leading order 
 ${\mathcal H}$ will depend on $\zeta_u$, $\zeta_v$ and $\rho$ because of the subtraction. The dependences will be canceled by those of TMD gluon density matrices and the soft factor. 
${\mathcal H}$ is obtained in \cite{MWZ}. But there are several typos and errors in constant terms. We will give here the corrected one. For NRQCD factorization we have made the expansion in the small velocity $v$, as discussed 
in the section of Introduction. We have only taken the leading order $v^0$.  However, the correction from the next-to-leading order of $v$, i.e., the relativistic correction, is at the same level of the importance as the ${\mathcal O}(\alpha_s)$-correction, as discussed in \cite{nrqcd}. Here we also include the relativistic correction. This correction can be extracted from the results in \cite{LMC}. 
We have: 
\begin{eqnarray}
{\mathcal H}(\zeta_u,\zeta_v, \rho) &=& 1 -\frac{4}{3} \frac{  \langle {\mathcal P}(^1S_0^{[1] } )\rangle  }{ m_Q^2 \langle  {\mathcal O} (^1 S_0^{[1]} )\rangle }   +  \frac{\alpha_s N_c}{4\pi}
 \left [  \ln^2 \frac{\zeta_u^2}{Q^2}+\ln^2\frac{\zeta_v^2}{Q^2} -\ln\rho^2\left (1+ 2 \ln\frac{\mu^2}{Q^2} \right )
  + 2 \ln\frac{\mu^2}{Q^2}  
\right.
\nonumber\\
  && \left.  + \frac{7}{2} \pi^2+\frac{2}{N_c^2} \biggr (
    5 -\frac{1}{4}\pi^2 \biggr )  \right ]   + {\mathcal O}(\alpha_s^2) +{\mathcal O}(v^4)
\label{HPART}
\end{eqnarray} 
with $Q^2 = 4 m_Q^2$. 
In the second term in Eq.(\ref{HPART}) there is a ratio of two NRQCD matrix elements defined in \cite{nrqcd}. 
This term is at order of $v^2$. The neglected correction from the expansion of the small velocity $v$ 
is now at order of $v^4$. 
\par   
After giving our main result in Eq.(\ref{FACSF}) it is worthy to discuss the problem of the so-called scheme-dependence in TMD factorization. The scheme-dependence 
arises because one can define different TMD parton distributions.   
In the case of TMD quark distributions one can define subtracted quark distributions to absorb the 
corresponding soft factor as suggested in \cite{JC11}. One may also use the definition from the soft-collinear effective theory
given in \cite{EIS}.  With different definitions one obtains the similar factorized result with different 
perturbative coefficients. The difference can be calculated perturbatively as discussed in \cite{JCR,TMDD}. 
Similarly, one can also work with different definitions of TMD gluon distributions. In this work, we only give our result in Eq.(\ref{FACSF}) with the unsubtracted TMD gluon distributions defined in Eq.(\ref{DEF}). Hence, the soft factor $\tilde S$ 
appears explicitly.  
\par 
The TMD gluon density matrix of $h_A$ or $h_B$ depends on the parameter $\zeta_u$ or $\zeta_v$, respectively. The dependence is determined by Collins-Soper equation. This equation can be used to resume terms of the large logarithms of $q_\perp/Q$ in perturbative coefficient functions of collinear factorization. In this way, one obtains the standard Collins-Soper-Sterman(CSS) resummation formalism in \cite{CSS}. However, in CSS formalism there can be certain freedom to re-define perturbative coefficient functions to make them process-independent or universal. This has been noticed in \cite{CFG}. An application by using the 
freedom is given in \cite{BCFG} for Higgs production in hadron collisions. Following the work in \cite{CFG}, the impact 
of the scheme-dependence in TMD factorization on the correspondingly derived resummation formalism has been studied in \cite{PSY}. The difference 
between different  schemes can be determined perturbatively. It is noted that our scheme of TMD factorization here and hence 
the corresponding resummation formalism are referred as Ji-Ma-Yuan scheme according to \cite{TMDD,PSY}. 

\par                     
From our result one can derive the resummation formula of $\ln(q_\perp/Q)$ for the production of $\eta_{c,b}$. The resummation for the production of $J/\psi$ or $\Upsilon$ has been studied in \cite{SYY}. In these resummations the quantum numbers of the produced heavy quark pair are fixed. The resummation for the production of a heavy quark pair in general case 
is studied in \cite{ZLLSY,CGT}.

\par
In the factorization formula in Eq.(\ref{FACSF}) the physical effects from initial hadrons in the processes 
only appear in TMD gluon density matrices. They take different parametrization 
forms for different hadrons, e.g., different spins of hadrons. With the formula one can derive the angular distribution 
for given hadrons in the initial state. But, the results can be very lengthy. Here, we consider a realistic case in which $h_B$ is of spin-0 or unpolarized, $h_A$ is of spin-1/2. 
The classification of the gluon TMD distributions of a spin-1/2 hadron is given in Eq.(\ref{Dec-M}). For $h_B$ we have:
\begin{eqnarray}
\Gamma_{B}^{\mu\nu}(k_B, \zeta_v^2 ) = -\frac{1}{2} g_\perp^{\mu\nu} \tilde f_g (y,k_{B\perp})
   + \frac{1}{2 M_B^2} \left (k_{B\perp}^\mu k_{B\perp}^\nu  +  \frac{1}{2} g_\perp^{\mu\nu} k_{B\perp}^2 \right ) \tilde H^\perp (y,k_{B\perp} ). 
\end{eqnarray} 
With the factorization formula we can derive the result of the differential cross-section in the considered case as:
\begin{eqnarray} 
  \frac{d \sigma (x,y, q_\perp) } {d x d y d^2 q_\perp  }  & = & \frac{2 \pi \delta (xy s- Q^2) }{ Q^2 } \sigma_0 {\mathcal H} 
    \biggr [  A (x,y, q_\perp)   + \tilde s_\perp \cdot 
    q_\perp B (x,y,q_\perp) \biggr ],
\label{dsigF}     
\end{eqnarray}        
where $\vec s_\perp$ 
is the transverse-spin vector and $\tilde s_\perp^\mu = \epsilon_\perp^{\mu\nu} s_{\perp\nu}$. 
The two coefficient functions are expressed with gluon TMD distributions 
as: 
\begin{eqnarray}
   A (x,y, q_\perp) &=&\frac{1}{2}  \int d^2 k_{A\perp} d^2 k_{B\perp} d^2 \ell_\perp \delta^2 (k_{A\perp} + k_{B\perp}+\ell_\perp  -q_\perp)   \tilde S(\ell_\perp,\rho )
\nonumber\\   
   && \biggr [ f_g (x,k_{A\perp}) \tilde f_g (y,k_{B\perp}) 
      -\frac{1}{4 M_A^2 M_B^2} \biggr ( 2 (k_{A\perp}\cdot k_{B\perp} )^2 - k_{A\perp}^2 k_{B\perp}^2 \biggr ) H^\perp (x,k_{A\perp}) \tilde H^\perp (y,k_{B\perp})   \biggr ],      
\nonumber\\
   B(x,y,q_\perp) &=&  \frac{-1}{2 M_A q_\perp^2} 
   \int d^2 k_{A\perp} d^2 k_{B\perp} d^2 \ell_\perp \delta^2 (k_{A\perp} + k_{B\perp}+\ell_\perp  -q_\perp)   \tilde S(\ell_\perp,\rho ) 
\nonumber\\   
  && \biggr \{ k_{A\perp}\cdot q_\perp \tilde f_g (y,k_{B\perp}) G_T (x, k_{A\perp}) 
 +\frac{1}{8  M_B^2} \biggr [  \biggr ( 2 k_{A\perp}\cdot k_{B\perp} q_\perp\cdot k_{B\perp} + k_{B\perp}^2 k_{A\perp} \cdot q_\perp \biggr)  
\nonumber\\  
  &&  \biggr ( \Delta H_T (x,k_{A\perp}) - \frac{k_{A\perp}^2}{2 M_A^2} \Delta H_T^{\perp} (x,k_{A\perp} ) \biggr )
\nonumber\\  
    && +  2 \frac{k_{A\perp}\cdot k_{B\perp}}{ M_A^2} \biggr ( q_\perp\cdot k_{A\perp} k_{A\perp}\cdot k_{B\perp} 
      + k_{A\perp}^2 k_{B\perp} \cdot q_\perp \biggr )   
      \Delta H_T^\perp (x,k_{A\perp})  \biggr ] \tilde H^\perp(y,k_{B\perp})  \biggr \}.         
\end{eqnarray} 
At tree-level, i.e., with ${\mathcal H}=1$ in Eq.(\ref{dsigF}), parts of the above expression has been derived before. 
The function $A$ is given in \cite{BoPi} and the first term in $B$ 
is derived in \cite{SchZh}. In \cite{MWZ} the factorization of the contribution with $f_g \tilde f_g$ is examined at one-loop 
level. In this work, we have examined the factorization of all contributions at one-loop with the general result given 
in Eq.({\ref{FACSF}). 
\par     
The obtained differential cross-section does not depend 
on the helicity of the polarized hadron. But it depends on the transverse spin. This dependence will give an Single transverse-Spin Asymmetry(SSA).  The same SSA has been studied with the twist-3 collinear factorization 
in \cite{SchZh}. In the kinematical region $\Lambda_{QCD}\ll q_\perp \ll Q$, both the factorizations apply, as shown for SSA in Drell-Yan processes in \cite{JQVY}, where SSA is only generated by Sivers quark distribution in TMD factorization. In our case SSA is not only generated by the gluonic Sivers function $G_T$, but also 
by other two T-odd TMD gluons distributions. Therefore, the relation between the two factorizations for SSA needs to carefully examined.    
\par

\par\vskip20pt

\noindent 
{\bf 5. Summary} 
\par 
We have studied TMD factorization for $\eta_Q$-production in hadron collisions at low transverse momenta. If the factorization holds,  one can use the production process to extract TMD gluon distributions from experiments.  We have explicitly shown that the factorization holds at one-loop level, in which all nonperturbative effects are factorized into TMD gluon density matrices and  a soft factor defined as the vacuum expectation value 
of  product of gauge links.  There is only one perturbative coefficient standing for all perturbative effects. This coefficient is determined at one-loop level and implemented with the relativistic correction of $\eta_Q$. 
With the result here, all TMD gluon distribution functions at leading power can  be safely extracted from experimental data.

\par 
In general the initial gluons 
from the initial hadrons are off-shell,  and they can be with nonphysical polarizations.  This makes the study of the TMD factorization more complicated than that in the case of on-shell gluons. However,  with the complication we can still  show that the factorization holds and is gauge-invariant. It is interesting to note that at one-loop level there exist contributions from the super-leading region from different sets of diagrams. But, they are cancelled in the sum.  
With our factorized result at one-loop, it is possible to show the factorization beyond one-loop for the studied 
process. One can also use the approach employed here  for examining TMD factorization of other processes involving TMD gluon distributions mentioned 
in the Introduction. 

\par

\par\vskip20pt
\noindent
{\bf Acknowledgments}
\par
The work of J.P. Ma is supported by National Nature
Science Foundation of P.R. China(No.11275244). The partial support from the CAS center for excellence in particle 
physics(CCEPP) is acknowledged.

\par

\end{document}